\documentclass[a4paper,11pt]{article}
\usepackage{jheppub} 
\usepackage{lineno}

\usepackage{physics}
\usepackage{subcaption}
\usepackage{slashed}
\usepackage{layouts}
\usepackage{dsfont}

\newcommand{\CSE}{C_{\text{CSE}}}
\newcommand{\B}{\mathcal{B}}
\newcommand{\Q}{\mathcal{Q}}
\renewcommand{\L}{\mathcal{L}}

\title{The Chiral Separation Effect from lattice QCD at the physical point}

\author[a]{B. B. Brandt,}
\author[a]{G. Endr\H{o}di,}
\author[a]{E. Garnacho-Velasco,}
\author[a]{and G. Mark\'{o}}

\affiliation[a]{Fakult\"{a}t f\"{u}r Physik, Universit\"{a}t Bielefeld,\\
Universit\"{a}tsstra{\ss}e 25, 33615 Bielefeld, Germany}

\emailAdd{brandt@physik.uni-bielefeld.de}
\emailAdd{endrodi@physik.uni-bielefeld.de}
\emailAdd{egarnacho@physik.uni-bielefeld.de}
\emailAdd{gmarko@physik.uni-bielefeld.de}

\abstract{
In this paper we study the Chiral Separation Effect by means of first-principles 
lattice QCD simulations. For the first time in the literature, we determine the 
continuum limit of the associated conductivity using 2+1 flavors of dynamical 
staggered quarks at physical masses. 
The results reveal a suppression of the conductivity in the confined phase and 
a gradual enhancement toward the perturbative value for high temperatures.
In addition to our dynamical setup, we also investigate the impact of the
quenched approximation on the conductivity, using both staggered and Wilson quarks.
Finally, we highlight the relevance of employing conserved vector and anomalous axial currents
in the lattice simulations.
}

\begin{document}
\maketitle
\flushbottom

\section{Introduction} \label{Introduction}

The non-trivial topological structure of the QCD vacuum manifests itself in a variety of fascinating aspects that can be studied theoretically and probed experimentally. In this context, anomalous transport phenomena have attracted the most attention recently. These effects appear due to the interplay between quantum anomalies and electromagnetic fields or vorticity, explaining their intimate relation to the topological nature of QCD 
and representing the impact of event-by-event CP-violation in QCD~\cite{Kharzeev:2007jp}. 
The prime example among these phenomena is the Chiral Magnetic Effect (CME): the generation of a vector current due to magnetic fields and a chiral imbalance~\cite{Fukushima:2008xe}. In the last decade, major experimental campaigns have been launched in order to measure the CME in different setups ranging from condensed matter systems~\cite{Li:2014bha} to heavy ion collision experiments~\cite{STAR:2013ksd,STAR:2014uiw,STAR:2021mii}. 

Another anomalous transport phenomenon is the Chiral Separation Effect (CSE): the generation of an axial current in a dense and magnetized system~\cite{Son:2004tq,Metlitski:2005pr}. Although highly analogous to the CME in its formulation, the equilibrium interpretation of the two effects turns out to be quite different. Together with the CME, it is expected to form the Chiral Magnetic Wave (CMW), a collective excitation emerging in a magnetized environment~\cite{Kharzeev:2010gd}, which is also the subject of intense experimental searches~\cite{Huang:2015oca}.
The key parameter, describing the strength of the effect is the conductivity coefficient $\CSE$, to be defined in detail below.

\begin{figure}[t]
    \centering
\includegraphics{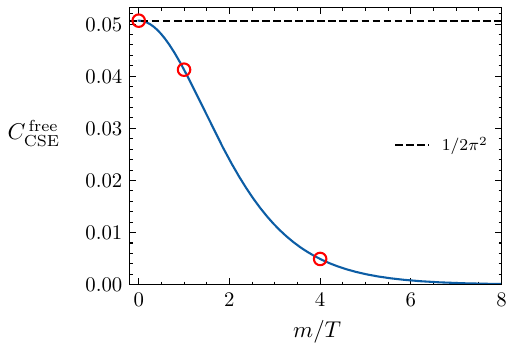}
    \caption{The CSE conductivity coefficient as a function of $m/T$ for non-interacting quarks. The marked points correspond to $m/T=0,1,4$, which will be the values we compare to in Sec.~\ref{sec:free quarks}.} 
        \label{fig:mTint}
\end{figure}

The analytical approaches to assess the CSE span a broad range and include for example chiral kinetic theory~\cite{Avdoshkin:2017cqp}, effective models of QCD~\cite{Gorbar:2011ya}, holography~\cite{Jimenez-Alba:2014pea} or the field correlator method~\cite{Zubkov:2023vvb}. The first calculations were performed for massless quarks in the absence of gluonic interactions, giving $\CSE=1/(2\pi^2)$~\cite{Son:2004tq}. It was later recognized that this value is not fixed by the axial anomaly but rather depends on the quark mass $m$ as well as the temperature $T$.
In fact, as we show in App.~\ref{sec:analytic appendix}, for non-interacting quarks, an analytical treatment of this problem gives
\begin{equation}
\label{eq:free tot}
     \CSE^{\text{\,free}} =
     \frac{1}{2\pi^2} \int_{0}^{\infty} \dd p \, \left[ 1+\cosh( \sqrt{p^2+(m/T)^2} ) \right]^{-1}\,,
\end{equation}
in agreement with ~\cite{Son:2004tq,Metlitski:2005pr}. 
The resulting curve is shown in Fig.~\ref{fig:mTint}, revealing a suppression of the conductivity coefficient for heavy quarks (or, equivalently, low temperatures), see also Refs.~\cite{Sheng:2017lfu,Lin:2018aon}. Note that this result corresponds to the linear behavior of the CSE current, valid for small chemical potentials, and this is the domain that we aim to study in this paper.

Given the sensitivity of the CSE to the mass and the temperature, it is natural to expect that the impact of gluonic interactions on $\CSE$ will also be substantial (see also Ref.~\cite{Gorbar:2013upa}). In the phenomenologically interesting region around the finite temperature QCD crossover~\cite{Aoki:2006we}, these interactions are non-perturbative. The most successful non-perturbative tool, based on the first principles of the underlying quantum field theory, is lattice QCD. However, at nonzero quark density -- which is required for the discussion of the CSE -- standard lattice simulations break down due to the complex action problem, see e.g.\ the recent review~\cite{Nagata:2021ugx}.
For this reason, many of the existing lattice investigations~\cite{Puhr:2016kzp,Buividovich:2020gnl} were not performed in full QCD, but either in the two-color version of it, or in the so-called quenched approximation -- both of which are free of the complex action problem. Even with this caveat in mind, some of the findings in the literature gave different predictions about whether QCD interactions at intermediate temperatures affect the value of $\CSE$~\cite{Buividovich:2020gnl} or not~\cite{Puhr:2016kzp}. The lattice regularization has also been employed to study the CSE for free fermions~\cite{Khaidukov:2017exf,Buividovich:2013hza} and with classical-statistical real-time simulations~\cite{Muller:2016jod,Mace:2016shq}.

In this work, we demonstrate for the first time how the CSE conductivity can be constructed using a Taylor-expansion around zero density, thereby enabling its treatment in full QCD. We determine $\CSE$ using dynamical staggered quarks with physical masses and carry out its continuum extrapolation for a broad range of temperatures in the transition region as well as deep in the confined phase. We shed further light on the impact of the quenched approximation and determine $\CSE$ in this setup using both staggered and Wilson quarks. In this setup, we also discuss the effect of increasing the quark masses away from the physical point. 
Finally, we also determine the CSE conductivity in the absence of gluonic interactions for both the staggered and the Wilson fermion discretizations -- this provides a useful benchmark of the appropriate lattice currents.

The present effort will be useful for studying further anomalous transport effects like the CME, for which most lattice simulations so far~\cite{Buividovich:2009wi,Buividovich:2009my,Yamamoto:2011ks,Buividovich:2013hza,Bali:2014vja,Astrakhantsev:2019zkr} were either based on indirect approaches or are yet to be performed in full QCD at the physical point.

\section{Conductivities and lattice methods} \label{Simulation setup}

As introduced above, the chiral separation effect amounts to the generation of an axial current $J_{\nu5}$ in the presence of a background magnetic field and nonzero quark density. The latter is parameterized by a chemical potential $\mu$ that couples to the corresponding conserved number density. The magnetic field $B$ is considered to be constant and homogeneous and, without loss of generality, pointing along the third spatial direction. Moreover, the magnetic field is measured in units of the elementary electric charge $e>0$, so that we can work with the renormalization group invariant combination $eB$. In this paper, we are interested in the leading-order behavior of the current for weak chemical potentials and weak magnetic fields. This response is described by the linear Taylor-coefficient $\CSE$ that we refer to as the CSE conductivity coefficient, or simply as CSE conductivity.

\subsection{Currents and chemical potentials}
\label{sec:curchempot}

For a single quark flavor, the definition of the axial current and the chemical potential is unambiguous. For several quark flavors, like in full QCD, there are different options which -- as we will find below -- give different results for the strength of the effect. One may consider the axial current
\begin{equation}
J_{\nu5}^i=\sum_f c^i_f\,\frac{T}{V}\int \dd^4x \, \bar\psi_f(x) \gamma_\nu \gamma_5 \psi_f(x)\,,
\label{eq:axcurdef}
\end{equation}
with different quantum numbers $i=\B,\Q$ or $\L$ in flavor space,
\begin{equation}
c_f^\B = \frac{1}{3}\,, \qquad
c_f^\Q = \frac{q_f}{e}\,,\qquad
c_f^\L =
\begin{cases}
1/3\,, & f=u,d\,, \\
0\,, & \textmd{otherwise}\,,\\
\end{cases}
\label{eq:cfBdef}
\end{equation}
where $f=u,d,s,\ldots$ labels the quark flavors and $q_f$ are the corresponding electric charges. The currents~\eqref{eq:axcurdef} are rendered intensive by dividing by the Euclidean four-volume $V/T$. The above choices correspond to the currents coupling to baryon number $\B$, electric charge $\Q$ or the so-called light baryon number $\L$~\cite{Borsanyi:2012cr}. The latter serves to mimic heavy-ion collision setups.

Analogously to the axial currents, the quark density can also be controlled either by a baryon chemical potential $\mu_\B$, an electric charge chemical potential $\mu_\Q$ or a light baryonic chemical potential $\mu_\L$, which couples to the corresponding vector currents,
\begin{equation}
    J_{\nu}^i=\sum_f c_f^i\,\frac{T}{V}\int \dd^4x \, \bar\psi_f(x) \gamma_\nu  \psi_f(x)\,.
\end{equation}
The chemical potentials for the individual quark flavors are $\mu_f=\mu_\B/3$ in the baryonic and $\mu_f=\mu_\Q \,q_f/e$ in the charge case, while they are set as $\mu_u=\mu_d=\mu_\L/3$, $\mu_f=0$ ($f\neq u,d$) for the light baryon chemical potential.

The most common choice in the flavor space corresponds to the quantum number $\Q$ for the current and $\B$ for the chemical potential. In this case, the CSE conductivity coefficient is defined via the leading order-behavior for the expectation value
\begin{equation}
    \label{eq:cse}
    \langle J^\Q_{35} \rangle=\CSE\, \mu_\B \, e B + \mathcal{O}(\mu_\B^3, B^3)\,.
\end{equation}
Here we use the short notation $\CSE \equiv C_{\textmd{CSE}}^{\Q\B}$ and we analogously define $C_{\textmd{CSE}}^{ij}$, where the first superscript $i\in\{\B,\Q,\L\}$ corresponds to the current and the second one $j\in\{\B,\Q,\L\}$ to the chemical potential. Note that the conductivity of Eq.~(\ref{eq:cse}) -- as well as the related conductivities $C_{\textmd{CSE}}^{ij}$ -- are defined through the leading-order response of the current to the chemical potential and the magnetic field at physical quark masses and are, generically, functions of the temperature. For free quarks at zero temperature, for instance, $\CSE$ vanishes, as discussed in the introduction.

Full lattice QCD simulations suffer from the complex action problem at $\mu_\B\neq0$, $\mu_\Q\neq0$ or $\mu_\L\neq0$. To circumvent this issue, we consider the Taylor-expansion of the current expectation value in the chemical potential. To leading order, this gives the first chemical potential derivative of the CSE current, which, using Eq.~(\ref{eq:cse}), yields
\begin{equation}
   \eval{\pdv{\langle J_{35}^\Q \rangle}{\mu_\B}}_{\mu_\B=0}= C_{\text{CSE}} \, e B\,,
   \label{eq:j35der}
\end{equation}
and analogously for the other flavor combinations. 
This, leading-order Taylor expansion of the current only requires simulations at zero density, free of the sign problem. Finally, $C_{\text{CSE}}$ can be extracted via numerical differentiation with respect to $eB$. We note that, alternatively, one could also perform the derivative with respect to the magnetic field analytically. On the lattice, such derivatives can be implemented using the methodology developed in Ref.~\cite{Bali:2015msa}, see also Refs.~\cite{Bali:2020bcn,Buividovich:2021fsa}. In the present case, this would lead to an expression including a three-point function with two vector currents and one axial current, revealing the relation between the conductivity and the anomalous triangle diagram.

For each choice of the flavor structures, there is an overall proportionality constant involving the quark baryon numbers or quark charges and the number of colors $N_c=3$, which can be factored out of the conductivity.
Since the coupling to the magnetic field always occurs via the quark charges, these overall factors read
\begin{align}
C_{\rm dof} &\equiv C_{\rm dof}^{\Q\B} = C_{\rm dof}^{\B\Q} = \frac{N_c}{3}\sum_f\left(\frac{q_f}{e}\right)^2, \quad\quad
C_{\rm dof}^{\Q\Q} = N_c\sum_f \left(\frac{q_f}{e}\right)^3, \nonumber \\
C_{\rm dof}^{\B\B} &= \frac{N_c}{9}\sum_f \frac{q_f}{e}\,, \qquad
C_{\rm dof}^{\Q\L} = \frac{N_c}{3}\sum_{f=u,d}\left(\frac{q_f}{e}\right)^2\,,
\qquad
C_{\rm dof}^{\B\L} = \frac{N_c}{9}\sum_{f=u,d}\frac{q_f}{e}\,,
\end{align}
and similarly for the remaining ones.
For massless, non-interacting quarks, these factors appear as overall normalizations in $\CSE$. (Note that $C_{\rm dof}^{\B\B}=0$ for three light quarks.) Note, however, that for massive, non-degenerate quarks each flavor contributes differently, depending on the choice of the quantum numbers. In the interacting case, disconnected diagrams are also different for $\B$, $\Q$ or $\L$. This implies that the individual $C_{\textmd{CSE}}^{ij}$ are not equivalent despite scaling out the above factors. Since these overall factors can always be restored, from this point we rescale all our results by $C_{\text{dof}}$ unless explicitly stated otherwise.

Throughout this paper, we are working in Euclidean space-time. Taking into account the relation between the Minkowski (indicated by the superscript $\rm M$) and Euclidean Dirac matrices ($\gamma_0^{\rm M}=\gamma_4$, $\gamma_i^{\rm M}=i\gamma_i$), we find that the Minkowski-space  observable reads
\begin{equation}
   \Re \eval{\pdv{\langle J_{35}^\Q\rangle}{\mu_\B}}_{\mu_\B=0}^{\rm M} = -\Im \eval{\pdv{\langle J_{35}^\Q\rangle}{\mu_\B}}_{\mu_\B=0} \,. 
   \label{eq:EuclMink}
\end{equation}
Therefore we consider, for the rest of the text, the imaginary part of the Euclidean two-point function calculated on the lattice. The real part of the Euclidean observable was checked to be consistent with zero.

\subsection{Lattice setup -- staggered quarks}

We begin by describing the setup with rooted staggered quarks, which we used to study the CSE in full dynamical QCD with up, down and strange quark flavors. Here, the partition function $\mathcal{Z}$ is written using the Euclidean path integral over the gluon links $U$ as
\begin{equation}
    \mathcal{Z}=\int \mathcal{D} U \exp[-\beta S_g] \,\prod_f \qty[ \det M_f(U,q_f,m_f)]^{1/4},
    \label{eq:partfunc}
\end{equation}
with $\beta=6/g^2$ denotes the inverse gauge coupling and $m_f$ the quark masses for each flavor $f=u,d,s$. In Eq.~\eqref{eq:partfunc}, $S_g$ is the gluonic action, which in our setup is the tree-level Symanzik action, and $M_f$ is the massive staggered Dirac operator with twice stout-smeared links. The Dirac operator contains the quark charges $q_u/2=-q_d=-q_s=e/3$. 
The quark masses are tuned to the physical point as a function of the lattice spacing $a$~\cite{Borsanyi:2010cj}. The simulations are carried out using four-dimensional lattice geometries with $N_s$ spatial and $N_t$ temporal points. The physical spatial volume is given by $V=L^3=(aN_s)^3$ and the temperature by $T=(aN_t)^{-1}$. The lattice sites are labeled by the four integers $n=\{n_1,n_2,n_3,n_4\}$ and $\hat\nu$ denotes the unit vector in the $\nu$ direction.
 
The magnetic field $B$ is included as a classical background field, so we do not consider dynamical photons in our setup. The electromagnetic potential enters the Dirac operator for the quark flavor $f$ in the same way as the $\mathrm{SU}(3)$ links, i.e.\ as a parallel transporter between two lattice points $u_{f\nu}=\exp(iaq_fA_\nu)$. These electromagnetic links are chosen such that they represent a homogeneous magnetic field in the $x_3$ direction. The flux of the field is quantized as $eB=6\pi N_b/(aN_s)^2$ with the integer flux quantum $N_b\in\mathds{Z}$~\cite{Bali:2011qj}.
Moreover, the chemical potential is also included in the exponential form, multiplying the temporal links by $\exp(\pm a\mu_f)$. Even though our simulations are at $\mu_f=0$, we will need this form to calculate the Taylor-expansion coefficients.

Within the staggered formalism, the quark fields $\psi_f$ are transformed in coordinate and Dirac space to the so-called staggered fields $\chi_f$ in order to partially diagonalize the Dirac operator. To proceed, we need to express the vector and axial currents in terms of $\chi_f$.
The prescription that maintains a conserved vector current and an anomalous (flavor-singlet) axial current are the point-split bilinears~\cite{Sharatchandra:1981si},
\begin{equation}
    J_\nu^f(n) = \Bar{\chi}_f(n)  \Gamma^f_\nu(n,m) \chi_f(m)\,, \qquad
        J^f_{\nu5}(n) = \Bar{\chi}_f(n)  \Gamma^f_{\nu5}(n,m) \chi_f(m)\,,
\end{equation}
involving the staggered counterparts of the Dirac matrices~\cite{Durr:2013gp},
  \begin{align}
  \label{eq:gammas}
    \Gamma^f_\nu(n,m)&=\dfrac{\eta_\nu(n)}{2}\left[U_\nu(n)u_{f\nu}(n) \,e^{a\mu_f \delta_{\nu4}}\delta_{n+\hat{\nu},m}+U^{\dagger}_\nu(n-\hat{\nu})u^{*}_{f\nu}(n-\hat{\nu}) \,e^{-a\mu_f \delta_{\nu4}}\delta_{n-\hat{\nu},m}\right], \nonumber \\
        \Gamma^f_{\nu5} &=\dfrac{1}{3!}\sum_{\rho,\alpha,\beta} \epsilon_{\nu\rho\alpha\beta}\,\Gamma^f_\rho\Gamma^f_\alpha\Gamma^f_\beta\,. 
\end{align}
Here, $\eta_\nu(n)=(-1)^{\sum_{\rho<\nu}n_\rho}$ are the staggered phases
and $\epsilon_{\nu\rho\alpha\beta}$ the totally antisymmetric four-index tensor with the convention
$\epsilon_{1234}=+1$. We mention that these Dirac structures depend explicitly on the links as well as on the magnetic field and the chemical potential. In particular, note that $\Gamma^f_{35}$ involves hoppings in the temporal direction and thus depends on $\mu_f$ (but not on the chemical potentials for the other quark flavors $\mu_{f'}$ with $f\neq f'$).

With these definitions, we can now perform the Grassmann integral over the staggered fields, giving the expectation value for the (volume-averaged) axial current,
\begin{equation}
 \langle J_{35}^{\Q} \rangle = 
 \frac{T}{V} \frac{1}{4}\sum_f \frac{q_f}{e}\left\langle \Tr \left(\Gamma^f_{35}M_f^{-1}\right)  \right\rangle\,,
\end{equation}
where $\Tr$ refers to a trace in color space and a summation over the lattice coordinates. The factor $1/4$ results from rooting.
Next, we can calculate the derivative~\eqref{eq:j35der} required for the CSE conductivity.
This derivative generates the usual disconnected and connected terms, together with an additional tadpole term arising due to the derivative of $\Gamma^f_{35}$ with respect to $\mu_\B$,
\begin{equation}
\begin{split}
   C_{\rm CSE} \,eB = \eval{\pdv{\langle J_{35}^\Q\rangle}{\mu_\B}}_{\mu_\B=0}=\dfrac{T}{V}\Bigg[\dfrac{1}{16}&\sum_{f,f'}c_f^\Q c_{f'}^\B\expval{\text{Tr}\qty(\Gamma^f_{35} M_f^{-1})\text{Tr}\qty(\Gamma^{f'}_{4} M_{f'}^{-1})}\\
    -\dfrac{1}{4} &\sum_f c_f^Q c_f^\B\expval{\text{Tr}\qty(\Gamma^f_{35} M_f^{-1}\Gamma^f_{4} M_f^{-1})}\\
    +\dfrac{1}{4}&\sum_f c_f^\Q c_f^\B\expval{\text{Tr}\qty(\dfrac{\partial\Gamma^f_{35}}{\partial \mu_f}M_f^{-1})}\Bigg]\,,
    \end{split} \label{eq:derstag}
\end{equation}
where the flavor coefficients~\eqref{eq:cfBdef} entered.
We emphasize that the tadpole term -- the last line in Eq.~\eqref{eq:derstag} -- appears due to the exponential form of introducing the chemical potential. This form is required to maintain gauge invariance on the lattice and to avoid chemical potential-dependent ultraviolet divergences~\cite{Hasenfratz:1983ba}.
The tadpole contribution is analogous to the one that appears in the calculation of quark number susceptibilities for staggered quarks, see e.g.\ Ref.~\cite{Endrodi:2011gv}. We also mention that in deriving~\eqref{eq:derstag} we exploited the charge conjugation symmetry of the $\mu_\B=0$ system, i.e.\ that the expectation value of the baryon density and that of the axial current vanish at $\mu_\B=0$ (we refer back to this point in Sec.~\ref{sec:quenched}).

The expectation values in Eq.~\eqref{eq:derstag} are to be evaluated at $\mu_\B=0$ but nonzero magnetic field $eB$. The conductivities $C_{\rm CSE}^{ij}$ for the other flavor quantum numbers can be calculated analogously, and differ from Eq.~\eqref{eq:derstag} by factors of $c_f^{i}c_{f'}^j$ in the disconnected and $c_f^ic_f^j$ in the connected and tadpole terms under the flavor sums.

To enable a direct comparison to existing lattice results in the literature, we also consider QCD in the quenched approximation. 
This amounts to dropping the fermion determinant from the partition function~\eqref{eq:partfunc}, while leaving the fermionic operators in the observables unchanged. This results in a non-unitary theory where quarks behave differently in operators as in loop diagrams. It is a commonly employed approximation that simplifies simulation algorithms significantly.
For completeness, in the quenched case we employ both the staggered and the Wilson discretization of fermions. 

Finally, we also calculate $\CSE$ in the absence of gluonic interactions, both for Wilson and for staggered quarks. To this end, we calculated the eigenvalues and eigenvectors of the staggered Dirac operator exactly and constructed the necessary traces from these (see App.~\ref{sec: free appendix}), while for Wilson fermions we relied on stochastic techniques to estimate the traces.

\subsection{Lattice setup -- Wilson quarks}

Next we describe our setup involving Wilson quarks. This discretization we only consider for free fermions and in the quenched approximation -- simulations with non-degenerate light Wilson quarks (differing in their electric charges) would be a computationally much more challenging setup. In this case, the currents are constructed from the bispinor fields $\psi_f$,
\begin{equation}
    J_\nu^f(n) = \Bar{\psi}_f(n)  \Gamma^f_\nu(n,m) \psi_f(m)\,, \qquad
        J^f_{\nu5}(n) = \Bar{\psi}_f(n)  \Gamma^f_{\nu5}(n,m) \psi_f(m)\,.
\end{equation}
Here the Dirac structure is the same as in the continuum, (apart from the Wilson term), but the conserved vector and anomalous axial currents again involve a point-splitting, just as for staggered quarks~\cite{Karsten:1980wd}
\begin{align*}
         \Gamma^f_\nu(n,m)&=\frac{1}{2}\Big[(\gamma_\nu-r)U_\nu(n) u_\nu(n) \,e^{a\mu_f\delta_{\nu4}} \delta_{m,n+\hat{\nu}} \\
         &\qquad\quad+(\gamma_\nu+r)U^\dagger_\nu(n-\hat{\nu})u^*_\nu(n-\hat{\nu})\,e^{-a\mu_f\delta_{\nu4}} \delta_{m,n-\hat{\nu}}\Big]\,,\\
            \Gamma^f_{\nu5}(n,m)&=\frac{1}{2}\Big[\gamma_\nu\gamma_5 U_\nu(n) u_\nu(n) \,e^{a\mu_f\delta_{\nu4}} \delta_{m,n+\hat{\nu}}\\
	&\qquad\quad+\gamma_\nu\gamma_5 U^\dagger_\nu(n-\hat{\nu}) u^*_\nu(n-\hat{\nu})\,e^{-a\mu_f\delta_{\nu4}}\delta_{m,n-\hat{\nu}}\Big]\,,
\end{align*}
where $r$ is the coefficient of the usual Wilson term in the Dirac operator (taken to be one in our setup). Notice the presence of the Wilson term in the vector current and its absence in the axial current\footnote{The former follows because gauge invariance requires the chemical potential to enter as parallel transporter in all hopping terms in the action, and the vector current can be written as the $\mu_f$-derivative of the action. The latter is quite nontrivial: the axial anomaly equation is satisfied by this choice because the terms proportional to $r$ (breaking the $\mathrm{U}_A(1)$ symmetry explicitly) approach exactly the topological term $\propto \epsilon_{\nu\rho\alpha\beta} G_{\nu\rho}G_{\alpha\beta}$ in the continuum limit, where $G_{\alpha\beta}$ is the gluon field strength tensor~\cite{Karsten:1980wd}.}.

Using the above definitions, the expectation value of the axial current reads,
\begin{equation} 
 \langle J_{35}^{\Q} \rangle = 
 \frac{T}{V} \sum_f \frac{q_f}{e}\left\langle \Tr \left(\Gamma^f_{35}M_f^{-1}\right)  \right\rangle\,.
\end{equation}
Since $\Gamma^f_{35}$ this time only involves hoppings in the $3$ direction, it is independent of the chemical potential. Thus, in comparison to the staggered case, our observable takes a simpler form,
\begin{equation}
\begin{split}
   C_{\rm CSE} \, eB = \eval{\pdv{\langle J^\Q_{35}\rangle}{\mu_\B}}_{\mu_\B=0}=\dfrac{T}{V}\Bigg[&\sum_{f,f'}c_f^\Q c_{f'}^\B\expval{\text{Tr}\qty(\Gamma^f_{35} M_f^{-1})\text{Tr}\qty(\Gamma^{f'}_{4} M_{f'}^{-1})}\\
    -&\sum_fc_f^\Q c_f^\B\expval{\text{Tr}\qty(\Gamma^f_{35} M_f^{-1}\Gamma^f_{4} M_f^{-1})} \Bigg]\,,
\end{split}
\label{eq:derwil}
\end{equation}
involving only disconnected and connected terms. The conductivities for the other flavor quantum numbers follow similarly.

In the literature, it is also customary to consider a local vector current
\begin{equation}
J_{\nu}^{f,\rm loc}(n) = \bar\psi_f(n) \gamma_\nu \psi_f(n)
\label{eq:localveccur}
\end{equation}
instead. Even though it is not conserved, it has the same quantum numbers as $J^f_\nu$ and is often employed in Wilson fermion simulations, such as for the study of various aspects of hadron physics at zero and non-zero temperature, for instance. It was also used to study the CME in quenched and dynamical QCD~\cite{Yamamoto:2011ks}. It is also possible to introduce a chemical potential $\mu_\B^{\rm loc}$ that couples to these local currents in the action.
The analogue of~\eqref{eq:derwil} now reads
\begin{equation}
\begin{split}
    C_{\rm CSE}^{\Q \B, \rm loc} \, eB = \eval{\pdv{\langle J^\Q_{35}\rangle}{\mu^{\rm loc}_\B}}_{\mu^{\rm loc}_\B=0}=\dfrac{T}{V}\Bigg[&\sum_{f,f'}c_f^\Q c_{f'}^\B\expval{\text{Tr}\qty(\Gamma^f_{35} M_f^{-1})\text{Tr}\qty(\gamma_4 M_{f'}^{-1})}\\
    -&\sum_fc_f^\Q c_f^\B\expval{\text{Tr}\qty(\Gamma^f_{35} M_f^{-1}\gamma_4 M_f^{-1})} \Bigg]\,.
\end{split}    
\label{eq:derwilloc}
\end{equation}

This introduction of a chemical potential is a naive extension of the continuum formulation to the discretized theory and it is well known that this leads to new ultraviolet divergences~\cite{Hasenfratz:1983ba}.
Therefore we use this setup with the non-conserved, local current merely for comparison, and we will test the impact of these ultraviolet divergences on the observable~\eqref{eq:derwilloc} below.

\section{Results} \label{Results}

To estimate the traces appearing in Eqs.~\eqref{eq:derstag}, \eqref{eq:derwil} and~\eqref{eq:derwilloc}, we use the standard noisy estimator technique. 

We performed our numerical simulations using $100$ Gaussian noise vectors, which was found to be sufficient to reliably calculate the observables. For free staggered fermions, we calculated the traces using the exact eigensystem instead, see App.~\ref{sec: free appendix} for details.

To determine $\CSE$, we calculated the current derivative for different magnetic fields, and extracted the coefficient from its slope with respect to $eB$. We considered a linear fit function with no offset term: a one parameter fit whose optimal value is found via a usual $\chi^2$ minimization method. The error analysis for each simulation is performed using the jackknife method with 10 bins, which is also used to propagate the error to the fit. This is what we refer to as the statistical error of $\CSE$. Furthermore, we repeat the fit, successively eliminating data points at the largest value of $eB$, until we are left with only one point. The largest difference in the slopes between the original fit and the different repetitions is what we consider the systematical error of $\CSE$. 

\begin{figure}[ht]
    \centering
   \includegraphics{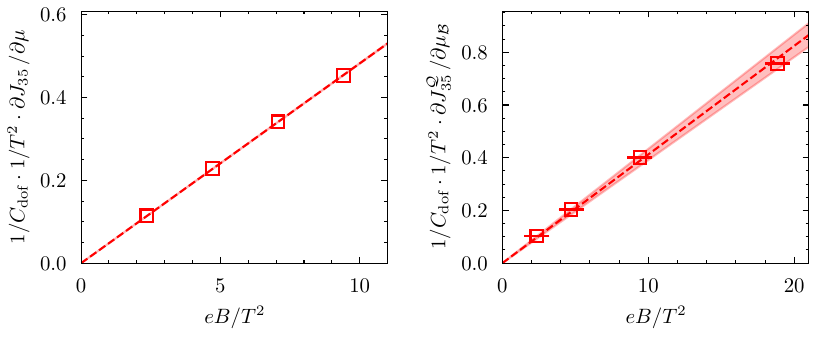}
          \caption{Derivative of the axial current with respect to the chemical potential as a function of the magnetic field for a $24^3 \times 6$ lattice with staggered quarks for free fermions (left) and in full QCD with $2+1$ flavors and physical quark masses at $T=305$ MeV (right). The dashed line represents the optimal fit, while the band is our estimate for its uncertainty, obtained by adding the statistical and systematical errors in quadrature.}
            \label{fig:fit}
\end{figure}

In Fig.~\ref{fig:fit} we show an example of the obtained results and the fits, both for free fermions and full QCD with staggered quarks. In these plots, we can see that the expected linear behavior is confirmed. Now we turn to the precise analysis of $\CSE$ in different situations.

\subsection{Free quarks}\label{sec:free quarks}
Since the free case is analytically solvable, we can use this simple situation as a check of our setup. 
We consider one color-singlet fermion with charge $q$ and mass $m$ -- this implies that the flavor quantum numbers are irrelevant and it is sufficient to treat a single chemical potential $\mu$ and axial current $J_{35}$. The overall proportionality constant is thus $C_{\rm dof}=(q/e)^2$, which is used to normalize the results.

\begin{figure}[ht]
    \centering
    \includegraphics{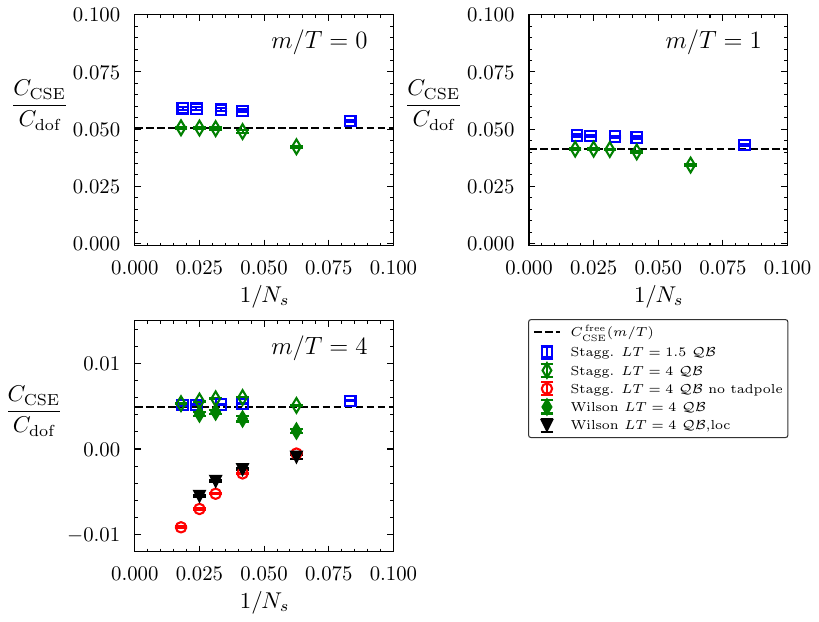}
    \caption{CSE conductivity coefficient for free fermions at different values of $m/T$ using the staggered discretization. For $m/T=4$, Wilson fermions as well as discretizations with a non-conserved vector current are also included. The dashed lines correspond to the marked values from Fig~\ref{fig:fit}.}
            \label{fig:free}
\end{figure}

In Fig.~\ref{fig:free} we show the results for free staggered fermions at different $m/T$ values. 
To compare to the analytic formula, both the continuum limit ($a\to0$), as well as the thermodynamic limit ($L\to\infty$) need to be taken.
The different panels of Fig.~\ref{fig:free} correspond to the continuum extrapolations for different aspect ratios $LT$ by increasing $N_s$ and $N_t$ (with $LT$ and $m/T$ kept constant). For small values of $m/T$, finite-size effects are expected to be sizeable, and the results confirm this expectation. In particular, $LT=4$ is already found to agree with the infinite volume limit for all masses.
A cross-check with Wilson fermions is also shown for the largest value of $m/T$.

The main message that we get from this calculation is that $\CSE$ approaches the value given by Eq.~(\ref{eq:free tot}) when the volume is sufficiently large and the continuum limit is taken. In the case of Wilson fermions, we also learn that the correct setup is to consider a conserved vector current and the anomalous axial current. If instead we use the local vector current~\eqref{eq:localveccur}, the analytical result is no longer recovered and the continuum limit shows a possibly divergent behavior. A very similar phenomenon occurs for staggered fermions if we exclude the tadpole term of~\eqref{eq:derstag}, again leading to a non-conserved (in fact ultraviolet divergent) vector current. Having cross-checked the consistency of our setup with the analytical result, we can proceed toward the physical setting by turning on the gluonic interactions.

\subsection{Quenched QCD}
\label{sec:quenched}

Before moving on to full QCD, there is an intermediate step where one can analyze the impact of gluons on the CSE: the quenched approximation. Furthermore, as we will see, the quenched theory reveals an interesting feature of the CSE due to the presence of an exact center symmetry. 
To appreciate this, next we briefly introduce the notion of the center symmetry and the Polyakov loop.

In the absence of dynamical fermions, the system undergoes a first-order phase transition from confined to deconfined matter at around $T_c^q\approx270$ MeV \cite{Boyd:1996bx}, for which the Polyakov loop
\begin{equation}
    P=\frac{1}{V}\sum_n \Tr \qty[\,\prod_{n_4=0}^{N_t-1} U_4(n)]\, ,
\end{equation}
acts as the order parameter.

This behavior arises from the $\mathrm{Z}(3)$ center symmetry of the gauge action, which is invariant under transformations $U_4(n)\rightarrow zU_4(n)$, with $z\in \mathrm{Z}(3)$, while the Polyakov loop is not, enforcing a vanishing expectation value of $P$. Below the transition temperature, center symmetry is intact and on typical configurations $P=0$. In turn, in the deconfined phase center symmetry is spontaneously broken, and the theory chooses one of the three vacua with $\arg P=0$, $\pm2\pi/3$ with equal probability.

\begin{figure}[ht]
    \centering
    \includegraphics{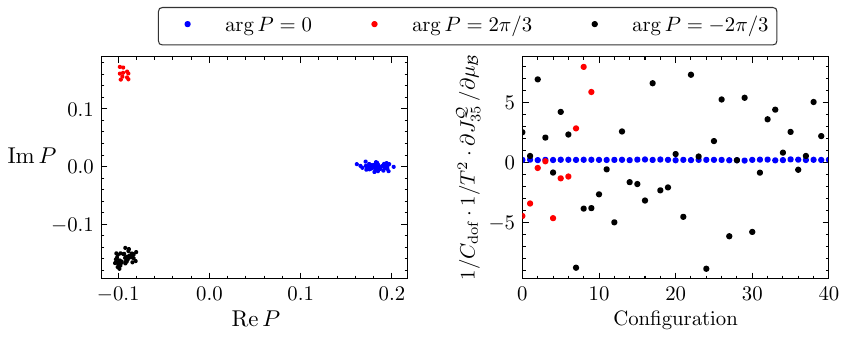}
    \caption{Results for the Polyakov loop (left) and the derivative of the current with staggered fermions (right) in an ensemble of 100 configurations on a $32^3\times8$ lattice at $T\approx 400$ MeV and at a magnetic field of $eB=0.74 \text{ GeV}^2$. The three sectors can be clearly distinguished at this temperature, and the contribution of configurations with imaginary Polyakov loops is observed to be very different as compared to those in the real sector.}
    \label{fig:poly}
\end{figure}

The Polyakov loop sectors at high temperature (where these are spatially approximately homogeneous) can be mimicked by an imaginary baryon chemical potential $i\mu_\B/T=0,\pm2\pi/3$. 

As we show in App.~\ref{sec:analytic appendix}, nonzero imaginary chemical potentials give a non-trivial contribution\footnote{In fact, at $i\mu_\B\neq0$, the expectation values of the baryon density and the axial current are nonzero, giving rise to an additional term $\propto\expval{\text{Tr}(\Gamma_{35}^fM_f^{-1})}\expval{\text{Tr}(\Gamma_4^{f'}M_{f'}^{-1})}$ in Eq.~\eqref{eq:derstag} for example. This term is included for the comparison in Fig.~\ref{fig:poly}.} to the observable~\eqref{eq:j35der}. In Fig.~\ref{fig:poly}, we present an example of this behavior at a temperature of $T\approx 400$ MeV. We observe that the Polyakov loop populates the three different, well separated sectors (left panel). In the right panel of Fig.~\ref{fig:poly}, we show the contributions of the individual sectors to the real part of the observable\footnote{That is to say, the imaginary part of the Euclidean observable, see Eq.~\eqref{eq:EuclMink}.}, indicating the effect of the effective imaginary baryon chemical potential. The fluctuations of the current derivative are observed to be enhanced drastically in the imaginary sectors.

In full QCD, the quarks explicitly break the $\mathrm{Z}(3)$ symmetry and, consequently, the theory is always in the real Polyakov loop sector -- corresponding to a vanishing imaginary baryon chemical potential. This indicates that the relevant QCD contribution to the CSE originates from configurations with real Polyakov loops. Therefore, for the quenched analysis of $\CSE$ below, we rotate all our gauge configurations to this sector by the appropriate center transformation. We note that while this procedure has a clear impact at high temperatures, where the Polyakov loop background is homogeneous (and thus equivalent to a nonzero $i\mu_B$), around $T_c^q$ the non-trivial spatial distribution of center clusters (see e.g.~\cite{Endrodi:2014yaa}) complicates the interpretation. This is a shortcoming of the quenched approximation for the CSE.

\begin{figure}[ht]
    \centering
    \includegraphics{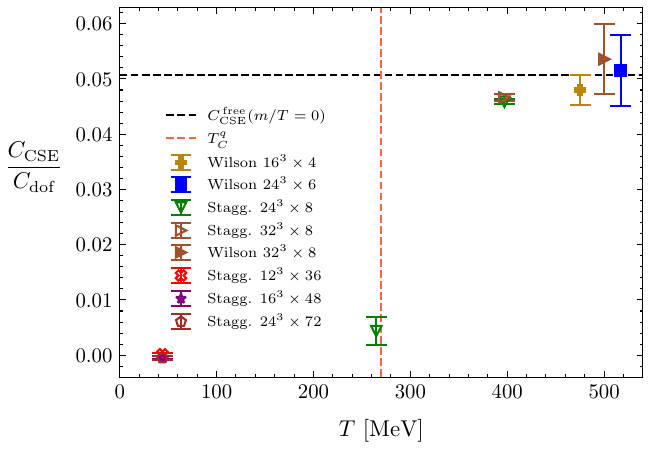}
    \caption{Results for $\CSE$ with Wilson and staggered fermions in the quenched approximation, measured on configurations generated using the plaquette gauge action. The configurations were rotated to the real Polyakov loop sector (for details see the text). The valence pion mass is set to $M_\pi\approx415$ MeV for staggered quarks and $M_\pi\approx 710$ MeV for Wilson quarks. The quenched critical temperature $T_c^q$ is indicated by the dashed vertical line.}
    \label{fig:qnch}
\end{figure}

Having this caveat in mind, 
we present the results for $\CSE$ in the quenched approximation in Fig.~\ref{fig:qnch}. We use configurations generated with the plaquette gauge action at $\beta=5.845, 5.9, 6.0, 6.2, 6.26$ and $6.47$, and use both staggered and Wilson fermions in the valence sector. 
We employed these gauge configurations already in Refs.~\cite{Bali:2017ian,Bali:2018sey}.
For the staggered discretization, the quark masses were tuned to a pion mass of $M_\pi\approx415$ MeV, while for Wilson fermions we use $M_\pi\approx710 $ MeV. The latter choice is motivated by the CME study~\cite{Yamamoto:2011ks}.

Although here we only present results at a few temperatures, we can clearly observe two distinct regimes: in the confined phase the CSE is severely suppressed, reaching zero toward $T\approx0$, while at temperatures well above $T_c^q$, the result approaches the massless free fermion value. One may compare this with the findings of the quenched study~\cite{Puhr:2016kzp}, where no corrections due to QCD interactions were found. However, in that case the quenched configurations were probed with a massless overlap Dirac operator in the valence sector, complicating the interpretation of the results. We emphasize moreover that according to our results, the transition between the regime where CSE is suppressed and the one where the massless case is approached, appears to occur in the vicinity of $T_c^q\approx 270\textmd{ MeV}$. All these hints given by the quenched results will be confirmed by the full QCD simulations, which we present next. 

\subsection{Full QCD at physical quark masses}

Finally, we present the main result of this study: the conductivity coefficient $C_{\text{CSE}}$ in full QCD, in particular for $N_f=2+1$ flavors of staggered fermions at physical quark masses. This is the first fully non-perturbative result for $C_{\text{CSE}}$ at the physical point.
The measurements were performed on an already existing ensemble of configurations for different magnetic fields~\cite{Bali:2011qj,Bali:2012zg}.
 
\begin{figure}[ht]
    \centering
    \includegraphics{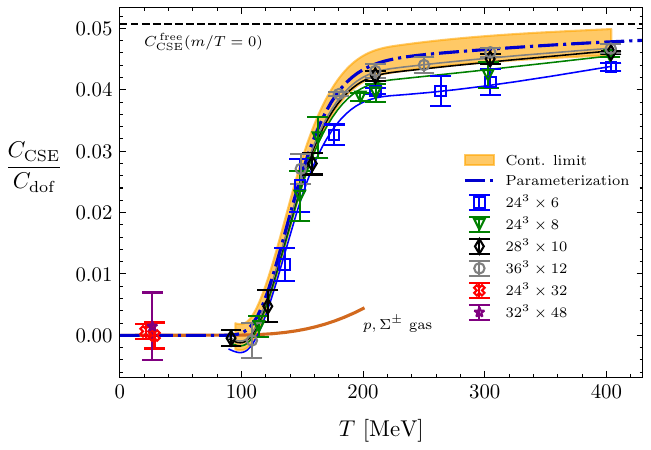}
    \caption{$\CSE$ for a broad range of temperatures in full QCD with $2+1$ flavors of staggered fermions. The continuum limit extracted via the spline fit procedure described in the text, together with its uncertainty, is indicated by the orange band. The result of our low-energy model involving $p$ and $\Sigma^{\pm}$ baryons is shown by the continuous line at low $T$, while the parameterization for $\CSE$ in the whole range of temperatures is displayed as a dashed-dotted line.}
    \label{fig:full_qcd}
\end{figure}

In Fig.~\ref{fig:full_qcd}, we present the dependence of the conductivity on the temperature using several finite-temperature lattice ensembles $24^3 \times 6$, $24^3 \times 8$, $28^3 \times 10$, $36^3 \times 12$ as well as two zero-temperature ensembles $24^3 \times 32$ and $32^3 \times 48$. The impact of the temperature on $\CSE$ is very pronounced, as we have already seen in the quenched results.
Above the QCD crossover temperature $T_c\approx 150\textmd{ MeV}$, $\CSE$ is found to approach the value corresponding to free massless quarks, in accordance with asymptotic freedom.
In turn, at temperatures below $T_c$, the conductivity decreases until reaching zero. This indicates a suppression of the CSE at low temperatures, which is consistent with a previous study in two-color QCD~\cite{Buividovich:2020gnl}.

In the low temperature region, we consider a simple model to describe the system: a non-interacting gas of hadronic degrees of freedom. For this particular observable, only electrically charged hadrons contribute that couple to chirality (i.e.\ the Dirac indices of $\gamma_5$)\footnote{We do not consider the impact of Wess-Zumino-Witten type terms~\cite{Scherer:2002tk} in our model.}. Thus, in our model we only consider a gas of protons and $\Sigma^{\pm}$ baryons. The contribution of heavier charged spin-$1/2$ baryons is negligible in the relevant temperature range, and we do not include spin-$3/2$ baryons either. The so constructed model also features a strong suppression of $\CSE$ in the confined regime and is found to agree with the lattice results for $T\lesssim 120\textmd{ MeV}$. It is also in qualitative agreement with other approaches to the CSE~\cite{Avdoshkin:2017cqp,Zubkov:2023vvb}. 

The continuum limit is carried out using the lattice data in the range of temperatures $90\textmd{ MeV}\lesssim T \lesssim 400$ MeV. To reliably extrapolate to the continuum\footnote{We note that the flavor-singlet axial vector current entering $\CSE$ is subject to multiplicative renormalization. The corresponding (perturbative) renormalization constant approaches unity in the continuum limit, both for Wilson and for staggered fermions~\cite{Constantinou:2016ieh,Bali:2021qem}. It might be used to reduce discretization errors, but in this work, we perform the continuum limit without including multiplicative renormalization constants.}, we consider a spline fit procedure combined with the continuum limit. For this, we consider a $T$-dependent spline fit of all lattice ensembles with $a$-dependent coefficients. The best fitting surface in the $a-T$ plane is found by minimizing $\chi^2/\text{dof}$. The details of the spline fit can be found in~\cite{Endrodi:2010ai}. The statistical error of this procedure is calculated using the jackknife samples, while we estimate the systematic error by repeating the spline fit removing the coarsest lattice. The maximum difference between the continuum limits obtained with these two data sets is taken as the systematic error, which is added in quadrature to the statistical one.

We further provide a complete parameterization of $\CSE$, to be used for comparisons with effective theories of QCD and in (anomalous) hydrodynamic descriptions of heavy-ion collisions. The parameterization, which is also shown in Fig.~\ref{fig:full_qcd}, smoothly connects the continuum extrapolation of the lattice results with the perturbative limit and the low-energy model. The details of the parameterization are given in App.~\ref{sec: param appendix}. In summary, we conclude that the CSE conductivity is an observable very sensitive to the finite temperature QCD crossover, and it may even be used to define its characteristic temperature. In particular, the inflection point of the parameterization is found at $T_c\approx 136(1)$ MeV. As an alternative definition, we also consider the temperature, where the conductivity assumes half its asymptotic value. This gives $T_c\approx 149(4)$ MeV.

\begin{figure}[ht]
    \centering
    \includegraphics{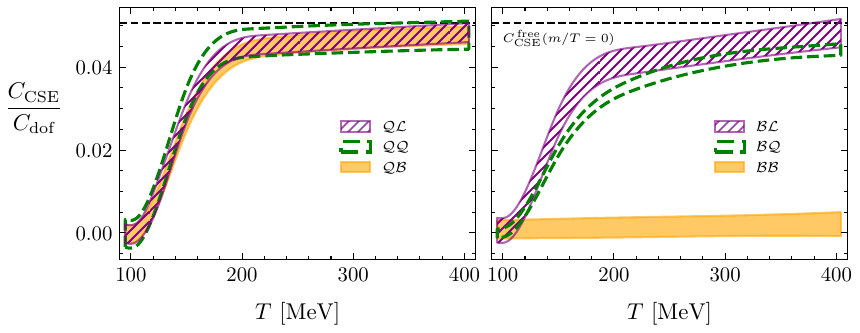}
    \caption{Continuum limits of $\CSE$ for different flavor quantum numbers (as defined in Eq.~\protect\eqref{eq:cfBdef}) for the axial current and the chemical potential. We present the phenomenologically relevant cases involving an electrically charged axial current (left panel) and a baryon axial current (right panel).}
    \label{fig:defs_comparison_QCD}
\end{figure}

The results above correspond to $\CSE=\CSE^{\Q\B}$, i.e.\ the response of the electrically charged axial current to the baryon chemical potential.
Next, we compare results for other combinations of the flavor quantum numbers $\Q,\B$ and $\L$. The corresponding proportionality factor $C_{\rm dof}$ is different for each (see Sec.~\ref{sec:curchempot}), but even after normalizing the conductivities by these factors, the results are not equivalent, since the different quark flavors interact with each other via gluons. Out of the nine possibilities, in Fig.~\ref{fig:defs_comparison_QCD} we present the continuum extrapolated results for the six combinations that are most interesting from a phenomenological point of view.

Except for the $\B\B$ case, all combinations show similar trends, being zero at low $T$ and approaching the massless free fermion result at high temperature. Still, differences at intermediate temperatures are visible, in particular in the cases with a baryonic axial current. The $\B\B$ setup is exceptional, because the three-flavor perturbative result in this case vanishes (see the discussion in Sec.~\ref{sec:curchempot}). Our full QCD result for $\CSE^{\B\B}$ also approaches zero for high temperatures. While at intermediate $T$, the perturbatively expected cancellation between the contributions of the three flavors is only partial, the result is still found to be much smaller than for the other quantum numbers.
Moreover, the data for $\CSE^{\B\B}$ is rather noisy due to the disconnected contribution in Eq.~\eqref{eq:derstag}, especially around $T_c$. While our current results do not enable us to resolve this temperature dependence, they do show that the continuum limit of this conductivity lies within the yellow band shown in the right panel of Fig.~\ref{fig:defs_comparison_QCD}. 

\section{Summary and outlook} \label{Summary and Outlook}
In this paper, we have presented a comprehensive study of the Chiral Separation Effect using first-principles lattice QCD simulations. In particular, we have calculated the conductivity coefficient $\CSE$ using dynamical staggered quarks with physical quark masses as well as quenched staggered and quenched Wilson quarks.
In addition, we calculated $\CSE$ in the absence of gluonic interactions both in the continuum and on the lattice with staggered and Wilson quarks.

In the free case, our analytic formula~\eqref{eq:free tot} reproduces the known results in the literature~\cite{Son:2004tq,Metlitski:2005pr}. Moreover, we find that both lattice discretizations reproduce the analytical result once the continuum limit is taken and the volume is sufficiently large. We highlight that the use of conserved vector currents and anomalous axial currents on the lattice is crucial -- non-conserved vector currents are demonstrated to lead to incorrect results.

Having cross-checked our setup, we moved on to full QCD, where we have determined the continuum limit of $\CSE$ for a broad range of temperatures at physical quark masses. At temperatures higher than $T_c$, the conductivity coefficient approaches the prediction for massless free fermions, as expected, while at temperatures below $T_c$, the conductivity goes to zero, revealing a strong suppression of the CSE in the confined phase. The result depends on the flavor quantum numbers used for the axial current and the chemical potential. Even after rescaling the conductivity by the multiplicative factor $C_{\rm dof}$ containing the degrees of freedom, for example a charge chemical potential induces slightly different baryonic or charge axial currents. 

A widely employed approximation to take into account gluonic interactions is quenched QCD, which we also explored in this paper. 
In the quenched theory, we highlighted the impact of the Polyakov loop sector on the CSE at high temperatures, an important subtlety that needs to be treated carefully when working in this approximation.
Only the real Polyakov loop sector delivers the physical result -- configurations in imaginary Polyakov loop sectors deviate from it and also come with drastically enhanced fluctuations. This issue can be overcome by rotating the configurations to the real sector by a center transformation. The results are in qualitative agreement with the picture based on our full QCD results: a strong suppression of the CSE at low temperatures and a tendency to the massless free fermion result at temperatures much higher than $T_c$.  

Our final results in full QCD are interpolated by a simple parameterization, explained in App.~\ref{sec: param appendix}, for use in hydrodynamic descriptions of heavy-ion collisions.
Finally, we note that our technique can be generalized to study the CME, which we plan to carry out in an upcoming publication. Performing the analogous simulations for that case will contribute to a better theoretical understanding of anomalous transport phenomena, as a counterpart to the large-scale experimental efforts being made to detect this effect.

\acknowledgments
This research was funded by the DFG (Collaborative Research Center CRC-TR 211 ``Strong-interaction matter under
extreme conditions'' - project number 315477589 - TRR 211) and by the Helmholtz Graduate School for Hadron and Ion Research (HGS-HIRe for FAIR). 
GE would like to express special thanks to the Mainz Institute for Theoretical Physics (MITP) of the Cluster of Excellence PRISMA+ (Project ID 39083149) for its hospitality and support.
The authors are moreover grateful for inspiring discussions with Pavel Buividovich, Kenji Fukushima, Dirk Rischke, S\"oren Schlichting, Igor Shovkovy and Lorenz von Smekal.

\appendix
\section{Analytical treatment of the CSE coefficient for free quarks}
\label{sec:analytic appendix}
We will now shortly summarize the continuum calculation leading to Eq.~\eqref{eq:free tot} for non-interacting fermions. 
As mentioned in the main text, it is sufficient to consider one colorless fermion flavor (with mass $m$ and charge $q$), a single chemical potential $\mu$ and axial current $J_{35}$. 
To enable a direct comparison to the lattice results, we will evaluate the $\mu$-derivative of the axial current in a homogeneous magnetic field background, in which case free fermion propagators are known exactly \cite{Schwinger:1951nm, Shovkovy:2012zn}. We will carry out the calculation starting out in Minkowski metric and extending to finite temperature during the calculation using the Matsubara formalism. Accordingly, the Dirac matrices are the Minkowski ones fulfilling $\{\gamma_\mu,\gamma_\nu\}=2\eta_{\mu\nu}=2\,\textmd{diag}(1,-1,-1,-1)$ contrary to the main text where Euclidean Dirac matrices are used.

Since the chemical potential, $\mu$ couples to the four-volume integral of the zeroth component of the vector current $J_\mu=\bar\psi\gamma_\mu\psi$, the $\mu$-derivative of the $z$-component of the chirality current can be written as
\begin{align}
    \left.\frac{\partial\langle J_{35}\rangle}{\partial\mu}\right|_{\mu=0}
    =  
    \int \dd^4x \int \dd^4y \,\langle \bar\psi(x)\gamma_3\gamma_5\psi(x)\bar\psi(y)\gamma_0\psi(y)\rangle\,.
    \label{eq:A1}
\end{align}
We regularize in the ultraviolet using the Pauli-Villars (PV) scheme, meaning that all diagrams are replicated by the regulator fields with coefficients $c_s$ and masses $m_s$. The details of the regularization is text-book knowledge found e.g.\ in Ref.~\cite{Itzykson:1980rh}. We recall here that we need three extra fields altogether, and reserving $s=0$ for the physical field (with physical mass $m$) the parameters will be
\begin{align}
    c_0&=c_1=1\,,\quad c_2=c_3=-1\,,\\
    m_0^2&=m^2\,,\quad m_1^2=m^2+2\Lambda^2\,,\quad m_2^2=m_3^2=m^2+\Lambda^2\,,
\end{align}
with $\Lambda\to\infty$ to be taken at the end of the calculation. From here on, the dependence on the regulator is made implicit and in fact, the Pauli-Villars fields turn out to give zero contribution to the CSE conductivity. However, their role is found to be crucial for the analogous calculation of the CME conductivity (which we will include in an upcoming publication).

The derivative according to Eq.~\eqref{eq:j35der} is proportional to the CSE coefficient, and using Wick's theorem we can rewrite the right hand side of Eq.~\eqref{eq:A1} to obtain
\begin{align}
    \label{eq:cse_PV_start}
    C_{\rm CSE}\, qB = \frac{iT}{V}\sum_{s=0}^{3} c_s\int \dd^4 x\int \dd^4y \Tr\left[\gamma_3\gamma_5 S_s(x,y) \gamma_0 S_s(y,x)\right]\,,
\end{align}
where the PV fields are already taken into account, and $S_s$ is the fermion propagator for the field $s$ in a homogeneous magnetic background. The latter reads~\cite{Shovkovy:2012zn}
\begin{align}
S_s(x,y) = \Phi(x,y)\int\frac{\dd^4p}{(2\pi)^4}{\rm \,e\,}^{-ip(x-y)}\widetilde S_s(p)\,.
\end{align}
Here 
\begin{align}
\Phi(x,y) = \exp\left[iqB(x_1+y_1)(x_2-y_2)\right]\,,
\end{align}
is the Schwinger phase, and
\begin{align}
\widetilde S_s(p) = \int_0^\infty \dd z {\rm \,e\,}^{iz(p_0^2 - m_s^2-p_3^2) - i\frac{p_1^2+p_2^2}{|qB|}\tan{(z|qB|)}}&\left[\slashed{p}+m_s+(p_1\gamma_2-p_2\gamma_1)\tan(zqB)\right]\nonumber\\
&\times\left[1-\gamma_1\gamma_2\tan(zqB)\right]\,.
\end{align}

To extend the calculation to finite temperature, we replace the frequency component of the integral by a sum over fermionic Matsubara frequencies
\begin{align}
    \int \frac{\dd p_0}{2\pi} \to T\sum_{n=-\infty}^\infty\,,
\end{align}
while also replacing $p_0\to i \omega_n= i 2\pi T(n+1/2)$ in $\widetilde S_s(p)$. Plugging this into \eqref{eq:cse_PV_start} and evaluating the trace over Dirac indices leads to
\begin{align}
    C_{\rm CSE} \,qB=4\sum_{s=0}^{3} c_s T\sum_{n}&\int\frac{\dd^3p}{(2\pi)^3}\int_0^\infty \dd z_1\, \dd z_2 {\rm \,e\,}^{i(z_1+z_2)(-\omega_n^2 - m_s^2-p_3^2)}\left(m_s^2-\omega_n^2+p_3^2\right) \nonumber\\
    &{\rm \,e\,}^{- i\frac{p_1^2+p_2^2}{qB}[\tan{(z_1qB)}+\tan{(z_2qB)}]}\left[\tan(z_1qB)+\tan(z_2qB)\right]\,.
\end{align}
The $p_1$ and $p_2$ integrals are simple Gaussians, and evaluating them factorizes the magnetic field dependence, uncovering an explicitly linear behavior in $B$. The rather simple formula emerges
\begin{align}
    C_{\rm CSE} =-\frac{1}{2\pi^2}\sum_{s=0}^{3} c_s T\sum_{n}&\int_{-\infty}^\infty \dd p_3\int_0^\infty \dd z_1\, \dd z_2 {\rm \,e\,}^{i(z_1+z_2)(-\omega_n^2 - m_s^2-p_3^2)}\left(m_s^2-\omega_n^2+p_3^2\right)\,,
\end{align}
allowing the evaluation of the $z_1$ and $z_2$ integrals. We find
\begin{align}
    C_{\rm CSE} =-\frac{1}{2\pi^2}\sum_{s=0}^{3} c_s T \sum_{n=-\infty}^\infty \int_{-\infty}^\infty \dd p_3\, \frac{m_s^2-\omega_n^2+p_3^2}{\left(\omega_n^2+p_3^2+m_s^2\right)^2}\,.
\end{align}

The Matsubara sum then evaluates to
\begin{align}
    C_{\rm CSE} =-\frac{1}{2\pi^2}\sum_{s=0}^{3} c_s \int_{-\infty}^\infty \dd p_3\, \frac{\dd n_F(E_p^{(s)})}{\dd E_p^{(s)}}\,,
\end{align}
where $n_F(x)=({\rm e}^{x/T}+1)^{-1}$ is the Fermi-Dirac distribution and $E_p^{(s)}=\sqrt{p_3^2+m_s^2}$. The derivative of the Fermi-Dirac distribution vanishes for infinite masses, therefore only the $s=0$, physical term remains from the sum over PV fields,
\begin{align}
    C_{\rm CSE} =-\frac{1}{\pi^2} \int_{0}^\infty \dd p_3\, \frac{\dd n_F(E_p)}{\dd E_p}\,,
\end{align}
which, after carrying out the derivative, gives Eq.~\eqref{eq:free tot} of the main text.

\begin{figure}[t]
    \centering
\includegraphics{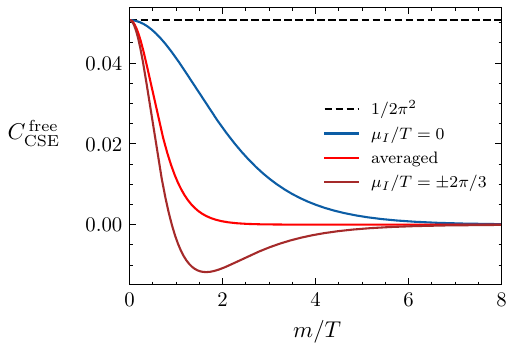}
    \caption{The CSE conductivity as a function of $m/T$ for non-interacting quarks, at different values of the imaginary chemical potential $\mu_I$. The averaged result corresponds to averaging over the three center sectors, that is $\mu_I/T=0,2\pi/3,-2\pi/3$.}
        \label{fig:mTint_mu}
\end{figure}

We point out here that generalizing this result to finite chemical potentials is rather simple and only amounts to replacing $n_F(E_p)$ by $[n_F(E_p+\mu)+n_F(E_p-\mu)]/2$. Specifically, e.g. for imaginary chemical potentials $\mu=i\mu_{I}$ we arrive at
\begin{align}
    C_{\rm CSE} =\frac{1}{2\pi^2} \int_{0}^\infty \dd p\, \frac{1+\cosh(\sqrt{p^2+(m/T)^2})\cos(\mu_I/T)}{\left[\cosh(\sqrt{p^2+(m/T)^2})+\cos(\mu_I/T)\right]^2}\,.
    \label{eq:CSEploop}
\end{align}
This formula reveals the impact of constant Polyakov loop backgrounds on the CSE conductivity. The three center sectors correspond to $\mu_I=\pm2\pi/3$ and $\mu_I=0$, for which~\eqref{eq:CSEploop} gives different responses, as can be seen in Fig.~\ref{fig:mTint_mu}. Interestingly, averaging over the three center sectors gives the same result as taking merely the real sector, except for a rescaling of the fermion mass,
\begin{equation}
\frac{1}{3}\sum_{k=0,1,2} C_{\rm CSE}(\mu_I/T=2\pi k/3, m/T) = C_{\rm CSE}(\mu_I=0,3m/T)\,.
\end{equation}
The averaged conductivity is also included in Fig.~\ref{fig:mTint_mu} for comparison. 

\section{CSE with free staggered fermions}
\label{sec: free appendix}
In this appendix, we illustrate how to calculate $\CSE$ for free staggered quarks using explicitly the eigenvalues and eigenvectors of the Dirac operator. Therefore, here we turn off the gluon fields and consider one color component only. Just as in App.~\ref{sec:analytic appendix}, we discuss a single quark flavor with mass $m$, one chemical potential $\mu$ and axial current $J_{35}$. Below we work on an $N_s^3\times N_t$ lattice and write everything in lattice units, setting $a=1$. In the non-interacting theory, the disconnected term of Eq.~\eqref{eq:derstag} vanishes, so we only need to calculate two terms
\begin{equation}
   \eval{\pdv{J_{35}}{\mu}}_{\mu=0}=\dfrac{1}{N_s^3N_t}\Bigg[-\dfrac{1}{4}\text{Tr}\qty(\Gamma_4 M^{-1}\Gamma_{35} M^{-1})
    +\dfrac{1}{4}\text{Tr}\qty(\dfrac{\partial\Gamma_{35}}{\partial \mu}M^{-1})\Bigg]\,,\label{eq:freederstag}
\end{equation}
and the expectation value indicating the fermion path integral was omitted for brevity.
Furthermore, for convenience we redefine the staggered phases as
\begin{equation}
    \eta_1(n)=1, \quad
    \eta_2(n)=(-1)^{n_1},\quad
    \eta_3(n)=(-1)^{n_1+n_2+n_4},\quad
    \eta_4(n)=(-1)^{n_1+n_2}\,.
\end{equation}

The massless staggered Dirac operator is antihermitian, $\slashed{D}^\dagger=-\slashed{D}$, therefore its eigenvalues are purely imaginary. Moreover, due to staggered chiral symmetry, $\{\slashed{D},\eta_5\}=0$ (with $\eta_5=(-1)^{n_1+n_2+n_3+n_4}$), the eigenvalues come in complex conjugate pairs,
\begin{equation}
    \slashed{D}\Psi_i=\pm i\lambda_i \Psi_i\,.
    \label{eq:B2}
\end{equation}
In addition, we will need the analogous eigensystem for $M M^\dagger = (\slashed{D}+m) (\slashed{D}+m)^\dagger = \slashed{D}\slashed{D}^\dagger+m^2$, so
\begin{equation}
    M M^\dagger\Psi_i=(\lambda_i^2+m^2) \Psi_i \,,
    \label{eq:B3}
\end{equation}
where each eigenvalue is doubly degenerate due to~\eqref{eq:B2}.
Using this eigensystem as basis, the traces in Eq.~\eqref{eq:freederstag} can be written as 
\begin{equation}
\begin{split}
   \eval{\pdv{J_{35}}{\mu}}_{\mu=0}=
    \dfrac{1}{N_s^3N_t}\Bigg[&-\dfrac{1}{4}\sum_{i,j}\dfrac{1}{(\lambda^2_i+m^2)(\lambda^2_j+m^2)}\Psi^\dagger_i \Gamma_4 M^\dagger \Psi_j\Psi^\dagger_j\Gamma_{35}M^\dagger \Psi_i\\
    &+\dfrac{1}{4}\sum_i\dfrac{1}{\lambda^2_i+m^2}\Psi^\dagger_i\dfrac{\partial\Gamma_{35}}{\partial \mu}M^\dagger\Psi_i\Bigg]\,,
    \end{split}
    \label{eq:spectralsum}
\end{equation}
where we inserted a complete set of eigenstates $\sum_j\Psi_j \Psi_j^\dagger=\mathds{1}$ in the first term and used $M^{-1}=M^\dagger(MM^\dagger)^{-1}$.

Next, we make use of the separability of the problem and reduce Eq.~\eqref{eq:B3} to one two-dimensional and two one-dimensional eigenproblems. This will enable us to determine the complete spectrum on much larger lattices than with a direct, four-dimensional approach. To this end, we write the free Dirac operator as $\slashed{D}=\slashed{D}_{12}+\slashed{D}_3+\slashed{D}_4$ with $\slashed{D}_{12}=\slashed{D}_1+\slashed{D}_2$ and
\begin{equation}
\slashed{D}_{\nu}(n,m)=\dfrac{\eta_\nu(n)}{2} \left[u_\nu(n)\delta_{n+\hat{\nu},m}-u^*_\nu(n-\hat{\nu})\delta_{n-\hat{\nu},m}\right]\equiv \eta_\nu(n)D_\nu(n,m)\,.
\end{equation}
Similarly, the staggered Dirac matrices~\eqref{eq:gammas} simplify to
\begin{align}
    \Gamma_\nu(n,m)=\dfrac{\eta_\nu(n)}{2} \left[u_\nu(n)\delta_{n+\hat{\nu},m}+u^*_\nu(n-\hat{\nu})\delta_{n-\hat{\nu},m}\right]\equiv \eta_\nu(n)S_\nu(n,m)\,.
\end{align}
Notice that the hop operators satisfy the property
\begin{equation}
    [S_1,S_3]= [S_1,S_4]=[S_2,S_3]=[S_2,S_4]=[S_3,S_4]=0\,,
    \label{eq:Scomm}
\end{equation}
because the $\mathrm{U}(1)$ links only enter in $S_1$ and $S_2$.
Moreover, 
\begin{equation}
\left.\frac{\partial S_4}{\partial \mu}\right|_{\mu=0}=D_4\,,
\label{eq:S4D4}
\end{equation}
due to the way that the chemical potential enters the temporal hoppings.

The squared operator $MM^\dagger=MM^\dagger_{12}+MM^\dagger_3+MM^\dagger_4$ separates into three terms that act in the respective subspaces and only depend on the respective coordinates -- thus they commute with each other.
Therefore, the eigenvectors in~\eqref{eq:B3} factorize,
\begin{equation}
    \Psi_{\{i_{12},i_3,i_4\}}(n_1,n_2,n_3,n_4)=\rho_{i_{12}}(n_1,n_2) \,\phi_{i_3}(n_3)\, \xi_{i_4}(n_4)\,,
    \label{eq:factmode}
\end{equation}
with $0\le i_{12}<N_s^2$, $0\le i_3<N_s$ and $0\le i_4<N_t$.\footnote{Note that the same separation does not hold for the eigensystem~\eqref{eq:B2}, since for example $[\slashed{D}_{12},\slashed{D}_3]\neq0$ due to the staggered phases. This is because $[MM^\dagger,\eta_5]=0$ but $[\slashed{D},\eta_5]\neq0$.} Below we use a shorthand notation and simply write $\Psi_i=\rho_i \phi_i \xi_i$. 

We proceed by expanding the operators appearing in the matrix elements in Eq.~\eqref{eq:spectralsum} into separate contributions that depend only on $n_{1,2}$, $n_3$ or $n_4$,
\begin{align}
    \Gamma_4&=
    (-1)^{n_1+n_2}S_4\,,\\
    M^\dagger&=-\slashed{D}_{12}-(-1)^{n_1+n_2+n_4} D_3-(-1)^{n_1+n_2}D_4+m\,,
\end{align}
and it can be shown with some algebra and Eq.~\eqref{eq:Scomm} that
\begin{equation}
    \Gamma_{35}= S_{12}S_4 (-1)^{n_2}\,,
\end{equation}
with $S_{12}=\{S_1,S_2\}/2$.
Finally, for the tadpole term we need the derivative of $\Gamma_{35}$ with respect to the chemical potential. The latter only appears in $S_4$. Therefore, using~\eqref{eq:S4D4},
\begin{align}
        \dfrac{\partial\Gamma_{35}}{\partial \mu}=S_{12}D_4(-1)^{n_2}\,.
\end{align}

The necessary products, appearing in~\eqref{eq:spectralsum} are therefore
\begin{equation}
\begin{split}
        \Gamma_4 M^\dagger=-&[(-1)^{n_1+n_2}\slashed{D}_{12}]_{12}\cdot[\mathds{1}]_3\cdot[S_4]_4\\
        -&[\mathds{1}]_{12}\cdot[D_3]_3\cdot[S_4(-1)^{n_4} ]_4\\
        -&[\mathds{1}]_{12}\cdot[\mathds{1}]_3\cdot[S_4D_4]_4\\
        +&m[(-1)^{n_1+n_2}]_{12}\cdot[\mathds{1}]_3\cdot[S_4]_4\,,
\end{split}
\end{equation}
where $[.]_{12}$, $[.]_3$ and $[.]_4$ indicate operators that only act in the respective spaces and only depend on the respective coordinates. Similarly, we obtain
\begin{equation}
\begin{split}
        \Gamma_{35} M^\dagger=-&[S_{12}(-1)^{n_2}\slashed{D}_{12}]_{12}\cdot[\mathds{1}]_3\cdot[S_4]_4\\
        -&[S_{12}(-1)^{n_1}]_{12}\cdot[D_3]_3\cdot[S_4(-1)^{n_4}]_4\\
        -&[S_{12}(-1)^{n_1}]_{12}\cdot[\mathds{1}]_3\cdot[S_4D_4]_4\\
        +&m[S_{12}(-1)^{n_2}]_{12}\cdot[\mathds{1}]_3\cdot[S_4]_4 \,,
\end{split}
\end{equation}
and 
\begin{equation}
\begin{split}
       \dfrac{\partial \Gamma_{35}}{\partial \mu} M^\dagger=&-[S_{12}(-1)^{n_2}\slashed{D}_{12}]_{12}\cdot[\mathds{1}]_3\cdot [D_4 ]_4\\
       &-[S_{12}(-1)^{n_1}]_{12}\cdot[D_3]_3\cdot [D_4 (-1)^{n_4}]_4\\
        &-[S_{12}(-1)^{n_1}]_{12}\cdot[\mathds{1}]_3\cdot[D_4 D_4]_4\\
        &+m[S_{12}(-1)^{n_2}]_{12}\cdot[\mathds{1}]_3\cdot[D_4 ]_4\,.
\end{split}
\label{eq:B18}
\end{equation}
Combining everything, Eq.~\eqref{eq:spectralsum} becomes
\begin{equation}
\begin{split}
          \dfrac{\partial J_{35}}{\partial \mu}\Big|_{\mu=0} =-&\dfrac{1}{N_s^3N_t}\dfrac{1}{4}\sum_{i,j}\dfrac{1}{(\lambda^2_i+m^2)(\lambda^2_j+m^2)}\\ &\Big\{[A_{ij}C_{ji}]_{12}\cdot[\delta_{ij}]_3\cdot[G_{ij}G_{ji}]_4+[A_{ij}E_{ji}]_{12}\cdot[\delta_{ij}]_3\cdot[G_{ij}H_{ji}]_4\\
          &+[A_{ij}E_{ji}]_{12}\cdot[\delta_{ij}J_{ji}]_3\cdot[G_{ij}I_{ji}]_4-m[A_{ij}F_{ji}]_{12}\cdot[\delta_{ij}]_3\cdot[G_{ij}G_{ji}]_4\\
          &+[\delta_{ij}C_{ji}]_{12}\cdot[\delta_{ij}]_3\cdot[H_{ij}G_{ji}]_4+[\delta_{ij}E_{ji}]_{12}\cdot[\delta_{ij}]_3\cdot[H_{ij}H_{ji}]_4\\
          &+[\delta_{ij}E_{ji}]_{12}\cdot[\delta_{ij}J_{ji}]_3\cdot[H_{ij}I_{ji}]_4-m[\delta_{ij}F_{ji}]_{12}\cdot[\delta_{ij}]_3\cdot[H_{ij}G_{ji}]_4\\
          &+[\delta_{ij}C_{ji}]_{12}\cdot[J_{ij}\delta_{ji}]_3\cdot[I_{ij}G_{ji}]_4+[\delta_{ij}E_{ji}]_{12}\cdot[J_{ij}\delta_{ji}]_3\cdot[I_{ij}H_{ji}]_4\\
          &+[\delta_{ij}E_{ji}]_{12}\cdot[J_{ij}J_{ji}]_3\cdot[I_{ij}I_{ji}]_4-m[\delta_{ij}F_{ji}]_{12}\cdot[J_{ij}\delta_{ij}]_3\cdot[I_{ij}G_{ji}]_4\\        
          &-m[B_{ij}C_{ji}]_{12}\cdot[\delta_{ij}]_3\cdot[G_{ij}G_{ji}]_4-m[B_{ij}E_{ji}]_{12}\cdot[\delta_{ij}]_3\cdot[G_{ij}H_{ji}]_4\\
          &-m[B_{ij}E_{ji}]_{12}\cdot[\delta_{ij}J_{ji}]_3\cdot[G_{ij}I_{ji}]_4+m^2[B_{ij}F_{ji}]_{12}\cdot[\delta_{ij}]_3\cdot[G_{ij}G_{ji}]_4\Big\}\\
          -&\dfrac{1}{N_s^3N_t}\dfrac{1}{4} \sum_i \dfrac{1}{\lambda^2_i+m^2}\\
          &\Big\{[C_{ii}]_{12}\cdot[\delta_{ii}]_3\cdot[K_{ii}]_4
          +[E_{ii}]_{12}\cdot[J_{ii}]_3\cdot[N_{ii}]_4\\
        &+[E_{ii}]_{12}\cdot[\delta_{ii}]_3\cdot[L_{ii}]_4
        -m[F_{ii}]_{12}\cdot[\delta_{ii}]_3\cdot[K_{ii}]_4 \Big\}\,,
     \end{split}
     \label{eq:freefinal}
\end{equation}
with
\begin{equation}
\begin{split}
    &A_{ij} \equiv \rho^\dagger_i(-1)^{n_1+n_2}\slashed{D}_{12} \rho_j\,, \\ 
    &C_{ij}\equiv \rho^\dagger_i S_{12}(-1)^{n_2}\slashed{D}_{12} \rho_j \,, \\
    &F_{ij}\equiv \rho^\dagger_i S_{12}(-1)^{n_2} \rho_j \,, \\
    &G_{ij}\equiv \xi^\dagger_i S_4\xi_j \,, \\
    &I_{ij}\equiv \xi^\dagger_i S_4(-1)^{n_4}\xi_j \,, \\
    &L_{ij}\equiv \xi^\dagger_i D_4 D_4\xi_j \,, 
\end{split}     
\qquad
\begin{split}
        &B_{ij} \equiv \rho^\dagger_i(-1)^{n_1+n_2} \rho_j \,,\\
        &E_{ij}\equiv \rho^\dagger_i S_{12}(-1)^{n_1} \rho_j \,, \\
        &J_{ij}\equiv \phi^\dagger_i D_3\phi_j \,,\\
        &H_{ij}\equiv \xi^\dagger_i S_4D_4\xi_j  \,, \\
        &K_{ij}\equiv \xi^\dagger_i D_4\xi_j \,, \\
        &N_{ij}\equiv \xi^\dagger_i D_4(-1)^{n_4}\xi_j \,.
\end{split}     
\label{eq:B20}
\end{equation}

We employ the LAPACK library to separately solve the two-dimensional eigenvalue problem in the $12$ plane in the presence of the magnetic field, as well as two one-dimensional problems in the $3$ and $4$ directions, both without electromagnetic phases. The eigensystem of the former problem gives Hofstadter's butterfly~\cite{Hofstadter:1976zz}, the spectrum of a well-known solid-state physics model. Its relevance for lattice QCD 
has been pointed out in Ref.~\cite{Endrodi:2014vza} and it has been generalized to the presence of gluonic interactions~\cite{Bruckmann:2017pft,Bignell:2020dze} as well as for inhomogeneous magnetic fields~\cite{Brandt:2023dir}.
The final value for the current derivative can be reconstructed using Eq.~\eqref{eq:freefinal}, and $\CSE$ can be extracted in the same way as explained in the main text by calculating the observable at different values of the magnetic field.

\section{Parameterization of the CSE conductivity coefficient}
\label{sec: param appendix}
In this appendix we provide a parameterization of the CSE conductivity in QCD as a function of temperature. We consider the continuum extrapolated data for $\CSE^{\Q\B}$ in Fig.~\ref{fig:full_qcd} and we smoothly interpolate to the prediction given by our model of a gas of protons and $\Sigma^\pm$ baryons at low $T$, while at high temperatures we impose an asymptotic approach to the result for a gas of free massless fermions ($m/T=0$ in Eq.~\eqref{eq:free tot}). We found the following parameterization to adequately describe the data, 
\begin{equation}
\label{eq:param}
    C_{\text{CSE}}(T)=\frac{1}{2\pi^2}\cdot \exp(-h/t^4)\cdot\frac{1+a_1/t+a_2/t^2+a_3/t^3}{1+a_4/t+a_5/t^2+a_6/t^3} \,, \qquad t=\frac{T}{1\text{ GeV}}\,,
\end{equation}
with the parameters included in Table.~\ref{tab:param}. As visible in Fig.~\ref{fig:full_qcd}, this function captures all details of our results -- more specifically, 
the confidence interval of the parameterization is about $95\%$.
\begin{table}[ht]
\centering
\begin{tabular}{|c|c|c|c|c|c|c|}
\hline
$h$     & $a_1$    & $a_2$  & $a_3$   & $a_4$   & $a_5$    & $a_6$   \\ \hline
0.000601 & -0.359512 &  0.049819 & -0.001234 & -0.346448 & 0.048787 & -0.002011 \\ \hline
\end{tabular}
\caption{Parameters of the function \eqref{eq:param}.}
\label{tab:param}
\end{table}

\bibliographystyle{JHEP}
\bibliography{biblio.bib}

\providecommand{\href}[2]{#2}\begingroup\raggedright\begin{thebibliography}{10}

\bibitem{Kharzeev:2007jp}
D.E.~Kharzeev, L.D.~McLerran and H.J.~Warringa, \emph{{The Effects of topological charge change in heavy ion collisions: 'Event by event P and CP violation'}}, \href{https://doi.org/10.1016/j.nuclphysa.2008.02.298}{\emph{Nucl. Phys. A} {\bfseries 803} (2008) 227} [\href{https://arxiv.org/abs/0711.0950}{{\ttfamily 0711.0950}}].

\bibitem{Fukushima:2008xe}
K.~Fukushima, D.E.~Kharzeev and H.J.~Warringa, \emph{{The Chiral Magnetic Effect}}, \href{https://doi.org/10.1103/PhysRevD.78.074033}{\emph{Phys. Rev. D} {\bfseries 78} (2008) 074033} [\href{https://arxiv.org/abs/0808.3382}{{\ttfamily 0808.3382}}].

\bibitem{Li:2014bha}
Q.~Li, D.E.~Kharzeev, C.~Zhang, Y.~Huang, I.~Pletikosic, A.V.~Fedorov et~al., \emph{{Observation of the chiral magnetic effect in ZrTe5}}, \href{https://doi.org/10.1038/nphys3648}{\emph{Nature Phys.} {\bfseries 12} (2016) 550} [\href{https://arxiv.org/abs/1412.6543}{{\ttfamily 1412.6543}}].

\bibitem{STAR:2013ksd}
{\scshape STAR} collaboration, \emph{{Fluctuations of charge separation perpendicular to the event plane and local parity violation in $\sqrt{s_{NN}}=200$ GeV Au+Au collisions at the BNL Relativistic Heavy Ion Collider}}, \href{https://doi.org/10.1103/PhysRevC.88.064911}{\emph{Phys. Rev. C} {\bfseries 88} (2013) 064911} [\href{https://arxiv.org/abs/1302.3802}{{\ttfamily 1302.3802}}].

\bibitem{STAR:2014uiw}
{\scshape STAR} collaboration, \emph{{Beam-energy dependence of charge separation along the magnetic field in Au+Au collisions at RHIC}}, \href{https://doi.org/10.1103/PhysRevLett.113.052302}{\emph{Phys. Rev. Lett.} {\bfseries 113} (2014) 052302} [\href{https://arxiv.org/abs/1404.1433}{{\ttfamily 1404.1433}}].

\bibitem{STAR:2021mii}
{\scshape STAR} collaboration, \emph{{Search for the chiral magnetic effect with isobar collisions at $\sqrt {s_{NN}}$=200 GeV by the STAR Collaboration at the BNL Relativistic Heavy Ion Collider}}, \href{https://doi.org/10.1103/PhysRevC.105.014901}{\emph{Phys. Rev. C} {\bfseries 105} (2022) 014901} [\href{https://arxiv.org/abs/2109.00131}{{\ttfamily 2109.00131}}].

\bibitem{Son:2004tq}
D.T.~Son and A.R.~Zhitnitsky, \emph{{Quantum anomalies in dense matter}}, \href{https://doi.org/10.1103/PhysRevD.70.074018}{\emph{Phys. Rev. D} {\bfseries 70} (2004) 074018} [\href{https://arxiv.org/abs/hep-ph/0405216}{{\ttfamily hep-ph/0405216}}].

\bibitem{Metlitski:2005pr}
M.A.~Metlitski and A.R.~Zhitnitsky, \emph{{Anomalous axion interactions and topological currents in dense matter}}, \href{https://doi.org/10.1103/PhysRevD.72.045011}{\emph{Phys. Rev. D} {\bfseries 72} (2005) 045011} [\href{https://arxiv.org/abs/hep-ph/0505072}{{\ttfamily hep-ph/0505072}}].

\bibitem{Kharzeev:2010gd}
D.E.~Kharzeev and H.-U.~Yee, \emph{{Chiral Magnetic Wave}}, \href{https://doi.org/10.1103/PhysRevD.83.085007}{\emph{Phys. Rev. D} {\bfseries 83} (2011) 085007} [\href{https://arxiv.org/abs/1012.6026}{{\ttfamily 1012.6026}}].

\bibitem{Huang:2015oca}
X.-G.~Huang, \emph{{Electromagnetic fields and anomalous transports in heavy-ion collisions --- A pedagogical review}}, \href{https://doi.org/10.1088/0034-4885/79/7/076302}{\emph{Rept. Prog. Phys.} {\bfseries 79} (2016) 076302} [\href{https://arxiv.org/abs/1509.04073}{{\ttfamily 1509.04073}}].

\bibitem{Avdoshkin:2017cqp}
A.~Avdoshkin, A.V.~Sadofyev and V.I.~Zakharov, \emph{{IR properties of chiral effects in pionic matter}}, \href{https://doi.org/10.1103/PhysRevD.97.085020}{\emph{Phys. Rev. D} {\bfseries 97} (2018) 085020} [\href{https://arxiv.org/abs/1712.01256}{{\ttfamily 1712.01256}}].

\bibitem{Gorbar:2011ya}
E.V.~Gorbar, V.A.~Miransky and I.A.~Shovkovy, \emph{{Normal ground state of dense relativistic matter in a magnetic field}}, \href{https://doi.org/10.1103/PhysRevD.83.085003}{\emph{Phys. Rev. D} {\bfseries 83} (2011) 085003} [\href{https://arxiv.org/abs/1101.4954}{{\ttfamily 1101.4954}}].

\bibitem{Jimenez-Alba:2014pea}
A.~Jimenez-Alba and L.~Melgar, \emph{{Anomalous Transport in Holographic Chiral Superfluids via Kubo Formulae}}, \href{https://doi.org/10.1007/JHEP10(2014)120}{\emph{JHEP} {\bfseries 10} (2014) 120} [\href{https://arxiv.org/abs/1404.2434}{{\ttfamily 1404.2434}}].

\bibitem{Zubkov:2023vvb}
M.A.~Zubkov and R.A.~Abramchuk, \emph{{Effect of interactions on the topological expression for the chiral separation effect}}, \href{https://doi.org/10.1103/PhysRevD.107.094021}{\emph{Phys. Rev. D} {\bfseries 107} (2023) 094021} [\href{https://arxiv.org/abs/2301.12261}{{\ttfamily 2301.12261}}].

\bibitem{Sheng:2017lfu}
X.-l.~Sheng, D.H.~Rischke, D.~Vasak and Q.~Wang, \emph{{Wigner functions for fermions in strong magnetic fields}}, \href{https://doi.org/10.1140/epja/i2018-12414-9}{\emph{Eur. Phys. J. A} {\bfseries 54} (2018) 21} [\href{https://arxiv.org/abs/1707.01388}{{\ttfamily 1707.01388}}].

\bibitem{Lin:2018aon}
S.~Lin and L.~Yang, \emph{{Mass correction to chiral vortical effect and chiral separation effect}}, \href{https://doi.org/10.1103/PhysRevD.98.114022}{\emph{Phys. Rev. D} {\bfseries 98} (2018) 114022} [\href{https://arxiv.org/abs/1810.02979}{{\ttfamily 1810.02979}}].

\bibitem{Gorbar:2013upa}
E.V.~Gorbar, V.A.~Miransky, I.A.~Shovkovy and X.~Wang, \emph{{Radiative corrections to chiral separation effect in QED}}, \href{https://doi.org/10.1103/PhysRevD.88.025025}{\emph{Phys. Rev. D} {\bfseries 88} (2013) 025025} [\href{https://arxiv.org/abs/1304.4606}{{\ttfamily 1304.4606}}].

\bibitem{Aoki:2006we}
Y.~Aoki, G.~Endr\H{o}di, Z.~Fodor, S.D.~Katz and K.K.~Szab\'o, \emph{{The Order of the quantum chromodynamics transition predicted by the standard model of particle physics}}, \href{https://doi.org/10.1038/nature05120}{\emph{Nature} {\bfseries 443} (2006) 675} [\href{https://arxiv.org/abs/hep-lat/0611014}{{\ttfamily hep-lat/0611014}}].

\bibitem{Nagata:2021ugx}
K.~Nagata, \emph{{Finite-density lattice QCD and sign problem: Current status and open problems}}, \href{https://doi.org/10.1016/j.ppnp.2022.103991}{\emph{Prog. Part. Nucl. Phys.} {\bfseries 127} (2022) 103991} [\href{https://arxiv.org/abs/2108.12423}{{\ttfamily 2108.12423}}].

\bibitem{Puhr:2016kzp}
M.~Puhr and P.V.~Buividovich, \emph{{Numerical Study of Nonperturbative Corrections to the Chiral Separation Effect in Quenched Finite-Density QCD}}, \href{https://doi.org/10.1103/PhysRevLett.118.192003}{\emph{Phys. Rev. Lett.} {\bfseries 118} (2017) 192003} [\href{https://arxiv.org/abs/1611.07263}{{\ttfamily 1611.07263}}].

\bibitem{Buividovich:2020gnl}
P.V.~Buividovich, D.~Smith and L.~von Smekal, \emph{{Numerical study of the chiral separation effect in two-color QCD at finite density}}, \href{https://doi.org/10.1103/PhysRevD.104.014511}{\emph{Phys. Rev. D} {\bfseries 104} (2021) 014511} [\href{https://arxiv.org/abs/2012.05184}{{\ttfamily 2012.05184}}].

\bibitem{Khaidukov:2017exf}
Z.V.~Khaidukov and M.A.~Zubkov, \emph{{Chiral Separation Effect in lattice regularization}}, \href{https://doi.org/10.1103/PhysRevD.95.074502}{\emph{Phys. Rev. D} {\bfseries 95} (2017) 074502} [\href{https://arxiv.org/abs/1701.03368}{{\ttfamily 1701.03368}}].

\bibitem{Buividovich:2013hza}
P.V.~Buividovich, \emph{{Anomalous transport with overlap fermions}}, \href{https://doi.org/10.1016/j.nuclphysa.2014.02.022}{\emph{Nucl. Phys. A} {\bfseries 925} (2014) 218} [\href{https://arxiv.org/abs/1312.1843}{{\ttfamily 1312.1843}}].

\bibitem{Muller:2016jod}
N.~M\"uller, S.~Schlichting and S.~Sharma, \emph{{Chiral magnetic effect and anomalous transport from real-time lattice simulations}}, \href{https://doi.org/10.1103/PhysRevLett.117.142301}{\emph{Phys. Rev. Lett.} {\bfseries 117} (2016) 142301} [\href{https://arxiv.org/abs/1606.00342}{{\ttfamily 1606.00342}}].

\bibitem{Mace:2016shq}
M.~Mace, N.~Mueller, S.~Schlichting and S.~Sharma, \emph{{Non-equilibrium study of the Chiral Magnetic Effect from real-time simulations with dynamical fermions}}, \href{https://doi.org/10.1103/PhysRevD.95.036023}{\emph{Phys. Rev. D} {\bfseries 95} (2017) 036023} [\href{https://arxiv.org/abs/1612.02477}{{\ttfamily 1612.02477}}].

\bibitem{Buividovich:2009wi}
P.V.~Buividovich, M.N.~Chernodub, E.V.~Luschevskaya and M.I.~Polikarpov, \emph{{Numerical evidence of chiral magnetic effect in lattice gauge theory}}, \href{https://doi.org/10.1103/PhysRevD.80.054503}{\emph{Phys. Rev. D} {\bfseries 80} (2009) 054503} [\href{https://arxiv.org/abs/0907.0494}{{\ttfamily 0907.0494}}].

\bibitem{Buividovich:2009my}
P.V.~Buividovich, M.N.~Chernodub, E.V.~Luschevskaya and M.I.~Polikarpov, \emph{{Quark electric dipole moment induced by magnetic field}}, \href{https://doi.org/10.1103/PhysRevD.81.036007}{\emph{Phys. Rev. D} {\bfseries 81} (2010) 036007} [\href{https://arxiv.org/abs/0909.2350}{{\ttfamily 0909.2350}}].

\bibitem{Yamamoto:2011ks}
A.~Yamamoto, \emph{{Lattice study of the chiral magnetic effect in a chirally imbalanced matter}}, \href{https://doi.org/10.1103/PhysRevD.84.114504}{\emph{Phys. Rev. D} {\bfseries 84} (2011) 114504} [\href{https://arxiv.org/abs/1111.4681}{{\ttfamily 1111.4681}}].

\bibitem{Bali:2014vja}
G.S.~Bali, F.~Bruckmann, G.~Endr\H{o}di, Z.~Fodor, S.D.~Katz and A.~Sch{\"a}fer, \emph{{Local CP-violation and electric charge separation by magnetic fields from lattice QCD}}, \href{https://doi.org/10.1007/JHEP04(2014)129}{\emph{JHEP} {\bfseries 04} (2014) 129} [\href{https://arxiv.org/abs/1401.4141}{{\ttfamily 1401.4141}}].

\bibitem{Astrakhantsev:2019zkr}
N.~Astrakhantsev, V.V.~Braguta, M.~D'Elia, A.Y.~Kotov, A.A.~Nikolaev and F.~Sanfilippo, \emph{{Lattice study of the electromagnetic conductivity of the quark-gluon plasma in an external magnetic field}}, \href{https://doi.org/10.1103/PhysRevD.102.054516}{\emph{Phys. Rev. D} {\bfseries 102} (2020) 054516} [\href{https://arxiv.org/abs/1910.08516}{{\ttfamily 1910.08516}}].

\bibitem{Borsanyi:2012cr}
S.~Bors\'anyi, G.~Endr\H{o}di, Z.~Fodor, S.D.~Katz, S.~Krieg, C.~Ratti et~al., \emph{{QCD equation of state at nonzero chemical potential: continuum results with physical quark masses at order $\mu^2$}}, \href{https://doi.org/10.1007/JHEP08(2012)053}{\emph{JHEP} {\bfseries 08} (2012) 053} [\href{https://arxiv.org/abs/1204.6710}{{\ttfamily 1204.6710}}].

\bibitem{Bali:2015msa}
G.~Bali and G.~Endr\H{o}di, \emph{{Hadronic vacuum polarization and muon g\ensuremath{-}2 from magnetic susceptibilities on the lattice}}, \href{https://doi.org/10.1103/PhysRevD.92.054506}{\emph{Phys. Rev. D} {\bfseries 92} (2015) 054506} [\href{https://arxiv.org/abs/1506.08638}{{\ttfamily 1506.08638}}].

\bibitem{Bali:2020bcn}
G.S.~Bali, G.~Endr\H{o}di and S.~Piemonte, \emph{{Magnetic susceptibility of QCD matter and its decomposition from the lattice}}, \href{https://doi.org/10.1007/JHEP07(2020)183}{\emph{JHEP} {\bfseries 07} (2020) 183} [\href{https://arxiv.org/abs/2004.08778}{{\ttfamily 2004.08778}}].

\bibitem{Buividovich:2021fsa}
P.V.~Buividovich, D.~Smith and L.~von Smekal, \emph{{Static magnetic susceptibility in finite-density $SU\left( 2\right) $ lattice gauge theory}}, \href{https://doi.org/10.1140/epja/s10050-021-00604-7}{\emph{Eur. Phys. J. A} {\bfseries 57} (2021) 293} [\href{https://arxiv.org/abs/2104.10012}{{\ttfamily 2104.10012}}].

\bibitem{Borsanyi:2010cj}
S.~Bors\'anyi, G.~Endr\H{o}di, Z.~Fodor, A.~Jakov\'ac, S.D.~Katz, S.~Krieg et~al., \emph{{The QCD equation of state with dynamical quarks}}, \href{https://doi.org/10.1007/JHEP11(2010)077}{\emph{JHEP} {\bfseries 11} (2010) 077} [\href{https://arxiv.org/abs/1007.2580}{{\ttfamily 1007.2580}}].

\bibitem{Bali:2011qj}
G.S.~Bali, F.~Bruckmann, G.~Endr\H{o}di, Z.~Fodor, S.D.~Katz, S.~Krieg et~al., \emph{{The QCD phase diagram for external magnetic fields}}, \href{https://doi.org/10.1007/JHEP02(2012)044}{\emph{JHEP} {\bfseries 02} (2012) 044} [\href{https://arxiv.org/abs/1111.4956}{{\ttfamily 1111.4956}}].

\bibitem{Sharatchandra:1981si}
H.S.~Sharatchandra, H.J.~Thun and P.~Weisz, \emph{{Susskind Fermions on a Euclidean Lattice}}, \href{https://doi.org/10.1016/0550-3213(81)90200-5}{\emph{Nucl. Phys. B} {\bfseries 192} (1981) 205}.

\bibitem{Durr:2013gp}
S.~D{\"u}rr, \emph{{Taste-split staggered actions: eigenvalues, chiralities and Symanzik improvement}}, \href{https://doi.org/10.1103/PhysRevD.87.114501}{\emph{Phys. Rev. D} {\bfseries 87} (2013) 114501} [\href{https://arxiv.org/abs/1302.0773}{{\ttfamily 1302.0773}}].

\bibitem{Hasenfratz:1983ba}
P.~Hasenfratz and F.~Karsch, \emph{{Chemical Potential on the Lattice}}, \href{https://doi.org/10.1016/0370-2693(83)91290-X}{\emph{Phys. Lett. B} {\bfseries 125} (1983) 308}.

\bibitem{Endrodi:2011gv}
G.~Endr\H{o}di, Z.~Fodor, S.D.~Katz and K.K.~Szab\'o, \emph{{The QCD phase diagram at nonzero quark density}}, \href{https://doi.org/10.1007/JHEP04(2011)001}{\emph{JHEP} {\bfseries 04} (2011) 001} [\href{https://arxiv.org/abs/1102.1356}{{\ttfamily 1102.1356}}].

\bibitem{Karsten:1980wd}
L.H.~Karsten and J.~Smit, \emph{{Lattice Fermions: Species Doubling, Chiral Invariance, and the Triangle Anomaly}}, \href{https://doi.org/10.1016/0550-3213(81)90549-6}{\emph{Nucl. Phys. B} {\bfseries 183} (1981) 103}.

\bibitem{Boyd:1996bx}
G.~Boyd, J.~Engels, F.~Karsch, E.~Laermann, C.~Legeland, M.~Lutgemeier et~al., \emph{{Thermodynamics of SU(3) lattice gauge theory}}, \href{https://doi.org/10.1016/0550-3213(96)00170-8}{\emph{Nucl. Phys. B} {\bfseries 469} (1996) 419} [\href{https://arxiv.org/abs/hep-lat/9602007}{{\ttfamily hep-lat/9602007}}].

\bibitem{Endrodi:2014yaa}
G.~Endr\H{o}di, C.~Gattringer and H.-P.~Schadler, \emph{{Fractality and other properties of center domains at finite temperature: SU(3) lattice gauge theory}}, \href{https://doi.org/10.1103/PhysRevD.89.054509}{\emph{Phys. Rev. D} {\bfseries 89} (2014) 054509} [\href{https://arxiv.org/abs/1401.7228}{{\ttfamily 1401.7228}}].

\bibitem{Bali:2017ian}
G.S.~Bali, B.B.~Brandt, G.~Endr\H{o}di and B.~Gl\"a\ss{}le, \emph{{Meson masses in electromagnetic fields with Wilson fermions}}, \href{https://doi.org/10.1103/PhysRevD.97.034505}{\emph{Phys. Rev. D} {\bfseries 97} (2018) 034505} [\href{https://arxiv.org/abs/1707.05600}{{\ttfamily 1707.05600}}].

\bibitem{Bali:2018sey}
G.S.~Bali, B.B.~Brandt, G.~Endr\H{o}di and B.~Gl\"a\ss{}le, \emph{{Weak decay of magnetized pions}}, \href{https://doi.org/10.1103/PhysRevLett.121.072001}{\emph{Phys. Rev. Lett.} {\bfseries 121} (2018) 072001} [\href{https://arxiv.org/abs/1805.10971}{{\ttfamily 1805.10971}}].

\bibitem{Bali:2012zg}
G.S.~Bali, F.~Bruckmann, G.~Endr\H{o}di, Z.~Fodor, S.D.~Katz and A.~Sch{\"a}fer, \emph{{QCD quark condensate in external magnetic fields}}, \href{https://doi.org/10.1103/PhysRevD.86.071502}{\emph{Phys. Rev. D} {\bfseries 86} (2012) 071502} [\href{https://arxiv.org/abs/1206.4205}{{\ttfamily 1206.4205}}].

\bibitem{Scherer:2002tk}
S.~Scherer, \emph{{Introduction to chiral perturbation theory}}, {\emph{Adv. Nucl. Phys.} {\bfseries 27} (2003) 277} [\href{https://arxiv.org/abs/hep-ph/0210398}{{\ttfamily hep-ph/0210398}}].

\bibitem{Constantinou:2016ieh}
M.~Constantinou, M.~Hadjiantonis, H.~Panagopoulos and G.~Spanoudes, \emph{{Singlet versus nonsinglet perturbative renormalization of fermion bilinears}}, \href{https://doi.org/10.1103/PhysRevD.94.114513}{\emph{Phys. Rev. D} {\bfseries 94} (2016) 114513} [\href{https://arxiv.org/abs/1610.06744}{{\ttfamily 1610.06744}}].

\bibitem{Bali:2021qem}
{\scshape RQCD} collaboration, \emph{{Masses and decay constants of the \ensuremath{\eta} and \ensuremath{\eta}' mesons from lattice QCD}}, \href{https://doi.org/10.1007/JHEP08(2021)137}{\emph{JHEP} {\bfseries 08} (2021) 137} [\href{https://arxiv.org/abs/2106.05398}{{\ttfamily 2106.05398}}].

\bibitem{Endrodi:2010ai}
G.~Endr\H{o}di, \emph{{Multidimensional spline integration of scattered data}}, \href{https://doi.org/10.1016/j.cpc.2011.03.009}{\emph{Comput. Phys. Commun.} {\bfseries 182} (2011) 1307} [\href{https://arxiv.org/abs/1010.2952}{{\ttfamily 1010.2952}}].

\bibitem{Schwinger:1951nm}
J.S.~Schwinger, \emph{{On gauge invariance and vacuum polarization}}, \href{https://doi.org/10.1103/PhysRev.82.664}{\emph{Phys. Rev.} {\bfseries 82} (1951) 664}.

\bibitem{Shovkovy:2012zn}
I.A.~Shovkovy, \emph{{Magnetic Catalysis: A Review}}, \href{https://doi.org/10.1007/978-3-642-37305-3_2}{\emph{Lect. Notes Phys.} {\bfseries 871} (2013) 13} [\href{https://arxiv.org/abs/1207.5081}{{\ttfamily 1207.5081}}].

\bibitem{Itzykson:1980rh}
C.~Itzykson and J.B.~Zuber, \emph{{Quantum Field Theory}}, International Series In Pure and Applied Physics, McGraw-Hill, New York (1980).

\bibitem{Hofstadter:1976zz}
D.R.~Hofstadter, \emph{{Energy levels and wave functions of Bloch electrons in rational and irrational magnetic fields}}, \href{https://doi.org/10.1103/PhysRevB.14.2239}{\emph{Phys. Rev. B} {\bfseries 14} (1976) 2239}.

\bibitem{Endrodi:2014vza}
G.~Endr\H{o}di, \emph{{QCD in magnetic fields: from Hofstadter's butterfly to the phase diagram}}, \href{https://doi.org/10.22323/1.214.0018}{\emph{PoS} {\bfseries LATTICE2014} (2014) 018} [\href{https://arxiv.org/abs/1410.8028}{{\ttfamily 1410.8028}}].

\bibitem{Bruckmann:2017pft}
F.~Bruckmann, G.~Endr\H{o}di, M.~Giordano, S.D.~Katz, T.G.~Kov\'acs, F.~Pittler et~al., \emph{{Landau levels in QCD}}, \href{https://doi.org/10.1103/PhysRevD.96.074506}{\emph{Phys. Rev. D} {\bfseries 96} (2017) 074506} [\href{https://arxiv.org/abs/1705.10210}{{\ttfamily 1705.10210}}].

\bibitem{Bignell:2020dze}
R.~Bignell, W.~Kamleh and D.~Leinweber, \emph{{Pion magnetic polarisability using the background field method}}, \href{https://doi.org/10.1016/j.physletb.2020.135853}{\emph{Phys. Lett. B} {\bfseries 811} (2020) 135853} [\href{https://arxiv.org/abs/2005.10453}{{\ttfamily 2005.10453}}].

\bibitem{Brandt:2023dir}
B.B.~Brandt, F.~Cuteri, G.~Endr\H{o}di, G.~Mark\'o, L.~Sandbote and A.D.M.~Valois, \emph{{Thermal QCD in a non-uniform magnetic background}},  \href{https://arxiv.org/abs/2305.19029}{{\ttfamily 2305.19029}}.

\end{thebibliography}\endgroup

\end{document}